\documentclass[12pt]{article}
\usepackage{comment}
\usepackage{amsmath}
\usepackage{amssymb}
\usepackage{graphicx}
\usepackage{here}
\usepackage{subcaption}
\usepackage{url}
\usepackage[sort&compress, numbers, merge]{natbib}

\usepackage{physics}
\usepackage[compat=1.1.0]{tikz-feynhand}
\usepackage{slashed}
\usepackage{mathtools}

\usepackage{todonotes}

\numberwithin{equation}{section}

\usepackage[colorlinks=true, linkcolor=blue, citecolor=blue,
urlcolor=black]{hyperref}

\begin{document}
\def\ps{\mathbf{p}}
\def\PS{\mathbf{P}}
\baselineskip 0.6cm
\def\simgt{\mathrel{\lower2.5pt\vbox{\lineskip=0pt\baselineskip=0pt
           \hbox{$>$}\hbox{$\sim$}}}}
\def\simlt{\mathrel{\lower2.5pt\vbox{\lineskip=0pt\baselineskip=0pt
           \hbox{$<$}\hbox{$\sim$}}}}
\def\simprop{\mathrel{\lower3.0pt\vbox{\lineskip=1.0pt\baselineskip=0pt
             \hbox{$\propto$}\hbox{$\sim$}}}}
\def\tr{\mathop{\rm tr}}
\def\SU{\mathop{\rm SU}}

\begin{titlepage}

\begin{flushright}
IPMU21-0025
\end{flushright}

\vskip 1.1cm

\begin{center}

{\Large \bf 
Muon $g-2$ in Gauge Mediation without SUSY CP Problem
}

\vskip 1.2cm
Masahiro Ibe$^{a,b}$, 
Shin Kobayashi$^{a}$, 
Yuhei Nakayama$^{a}$ and
Satoshi Shirai$^{b}$
\vskip 0.5cm

{\it

$^a$ {ICRR, The University of Tokyo, Kashiwa, Chiba 277-8582, Japan}

$^b$ {Kavli Institute for the Physics and Mathematics of the Universe
 (WPI), \\The University of Tokyo Institutes for Advanced Study, \\ The
 University of Tokyo, Kashiwa 277-8583, Japan}
}

\vskip 1.0cm

\abstract{
We discuss gauge mediated supersymmetry breaking models which explain the observed muon anomalous magnetic moment and the Higgs boson mass simultaneously.
The successful explanation requires 
the messenger sector which violates the 
relation motivated by the grand unification theory (GUT). 
The naive violation of the GUT relation, however, ends up with the CP problem.
We propose a model in which the phases of the gaugino masses are aligned despite the violation of the GUT relation.
We also consider a model which 
generates the $\mu$-term
and 
the additional Higgs soft masses squared without causing CP violation.
As a result, we find a  successful model which explains the muon anomalous magnetic moment and the Higgs boson mass.
The model is also free from 
 the CP, flavor-changing neutral current and the lepton flavor violation problems caused by the subdominant gravity mediation effects.
The lightest supersymmetric particles are gravitino/goldstini and the next-to-lightest ones are the Wino/Higgsinos in the typical parameter space.
We also study the LHC constraints. 
}

\end{center}
\end{titlepage}

\section{Introduction}
The Muon $g-2$ experiment at Fermilab reported the first results on the measurement of the muon anomalous magnetic moment, $a_\mu = (g-2)_\mu/2$.
The reported value of the combined result of Muon $g-2$ experiment at Fermilab and Brookhaven National Laboratory is
\begin{align}
\label{eq:deviation}
    a_\mu^{\mathrm{exp}} - a_\mu^{\mathrm{SM}} = (25.1 \pm 5.9) \times 10^{-10}\ ,
\end{align}
which corresponds to the 4.2\,$\sigma$ deviation from the Standard Model (SM) prediction based on the latest assessment of contributions from quantum electrodynamics (QED) up to the tenth order~\cite{Aoyama:2012wk,Aoyama:2019ryr}, vacuum polarization of hadrons~\cite{Davier:2017zfy,Keshavarzi:2018mgv,Colangelo:2018mtw,Hoferichter:2019gzf,Davier:2019can,Keshavarzi:2019abf,Kurz:2014wya}, light-by-light of hadrons~\cite{Melnikov:2003xd,Masjuan:2017tvw,Colangelo:2017fiz,Hoferichter:2018kwz,Gerardin:2019vio,Bijnens:2019ghy,Colangelo:2019uex,Pauk:2014rta,Danilkin:2016hnh,Jegerlehner:2017gek,Knecht:2018sci,Eichmann:2019bqf,Roig:2019reh,Blum:2019ugy,Colangelo:2014qya}, and electroweak processes~\cite{Jackiw:1972jz,Bars:1972pe,Fujikawa:1972fe,Czarnecki:2002nt,Gnendiger:2013pva} (See also Ref.~\cite{Aoyama:2020ynm} and references therein).
The deviation strongly indicates the physics beyond the SM, although higher statistical significance and the further refinement of the SM prediction are required to be conclusive.

Among various candidates for physics beyond the SM, which can explain the discrepancy, the minimal supersymmetric (SUSY) SM (MSSM) has been the most attractive one.
In the MSSM, the discrepancy of $a_\mu$ can be resolved when the masses of the sleptons and neutralinos/charginos are in the range of $\order{100}$\,GeV~\cite{Lopez:1993vi,Chattopadhyay:1995ae,Moroi:1995yh}.
In the resolution by the MSSM contribution, however, there are several concerns.
The colored SUSY particles in the $\order{100}$\,GeV range have been severely constrained by the results of the LHC searches~\cite{ATL-PHYS-PUB-2021-007}.
The light SUSY particles are also in tension with the observed Higgs boson mass~\cite{Okada:1990vk,Ellis:1990nz,Haber:1990aw}.
Besides, the light SUSY particles generically lead to large flavor changing neutral current (FCNC) effects, the lepton flavor violations (LFV), and CP violations.
In particular, 
there are correlations between the LFV/CP problems and the size of
the SUSY contribution, $a_\mu|_\mathrm{SUSY}$~\cite{Feng:2001sq,Calibbi:2006nq,Giudice:2012ms,Calibbi:2014yha}.

Given these concerns, it is interesting to discuss whether the gauge mediated SUSY breaking (GMSB) models~\cite{Dine:1981za,*Dimopoulos:1981au,*Dine:1981gu,*Dine:1982qj,*Dine:1982zb,*Nappi:1982hm,*AlvarezGaume:1981wy,*Dimopoulos:1982gm,Dine:1993yw,*Dine:1994vc,*Dine:1995ag} explain the discrepancy of $a_\mu$ consistently.
As an advantage of GMSB, it predicts 
flavor universal soft SUSY breaking parameters, which suppress the SUSY FCNC/LFV effects.
In GMSB, however, there are correlations between the squark masses and the slepton masses
when the messenger sector satisfies 
relations motivated by the grand unified theory (GUT).
Accordingly, in typical GMSB models, the squarks turn out to be too light to explain the Higgs boson mass when the sleptons are light enough to explain $a_\mu$.
Thus, the explanation of $a_\mu$ requires more extended GMSB models, for example, in which the SUSY spectrum deviates from the GUT relation.
Such extensions of the GMSB models often
lead to new sources of the CP violation,
which could ruin the successful features of the GMSB models.

In addition, the GMSB models have the so-called 
$\mu$ and $B$ problems. 
We need a mechanism to generate 
the Higgsino mass term and the holomorphic soft SUSY breaking mass parameter to achieve successful electroweak symmetry breaking.
In general, the mechanism to generate the $\mu$ and $B$ terms also induces additional CP violation phases.

In this paper, we discuss extended GMSB which violates the GUT relation
without CP violation.
We also consider the mechanism to generate the $\mu$-term developed in Refs.~\cite{Kitano:2006wm,Kitano:2006wz,Ibe:2006rc,Ibe:2007km},
which is also free from the CP problem.
As a result, we find that the extended GMSB can explain $a_\mu$ and the observed Higgs boson mass.
We also discuss the LHC constraints
on the SUSY spectra which explain $a_\mu$ and the Higgs boson mass.

The organization of the paper is as follows.
In Sec.~\ref{sec:GMSBvsGUT},
we discuss the necessity of the violation of the GUT relation in the messenger sector.
In Sec.~\ref{sec:Aligned},
we discuss a model which evades 
the relative phases of the gaugino masses.
In Sec.~\ref{sec:SSS}, we discuss the origin of the $\mu$-term.
We also discuss the SUSY CP and LFV problems due to the subdominant gravity mediation.
The LHC signatures are discussed in Sec.~\ref{sec:LHC}.
The final section is devoted to our conclusions.

\section{GMSB and CP Violation}
\label{sec:GMSBvsGUT}
\subsection{GMSB with GUT relation}
Let us first review the SUSY contribution to $a_\mu$ in the GMSB. 
In the minimal setup, 
the messenger chiral multiplets $(\Phi,\bar{\Phi})$ are in the $\mathbf{5}+\bar{\mathbf{5}}$  representation of
the minimal 
GUT gauge group, SU$(5)$. 
The messenger multiplets couple to the SUSY breaking field $Z$ via the superpotential,
\begin{align}
\label{eq:messenger}
W = k_D Z \bar{\Phi}_D\Phi_{\bar{D}}+ 
k_L Z\bar{\Phi}_{\bar{L}}\Phi_L \ ,
\end{align}
where $k_{D,L}$ are  coupling constants.
We decompose the messenger multiplets into $\Phi = (\Phi_{\bar{D}}, \Phi_L)$ and $\bar{\Phi} = (\bar{\Phi}_{D}, \bar{\Phi}_{\bar{L}})$ in accordance with the MSSM gauge charges.
To maintain the successful coupling unification, we require $k_{D}\sim k_L \sim k$ at the messenger scale.

For a while, we treat the SUSY breaking field as a spurious chiral supermultiplet which breaks supersymmetry with the vacuum expectation value (VEV),
\begin{align}
    \langle Z \rangle = A_Z + F_Z\theta^2\ . 
\end{align}
By using the phase rotation of $Z$ and superspace coordinate, $\theta$, i.e., 
$R$-symmetry rotation, we take $A_Z$ and $F_Z$ real positive.
We also take $k_{D,L}$ real positive by the phase rotation of the messenger multiplets.
Thus, in the minimal setup, there is no source of the CP violation.

In the minimal setup, the gaugino masses at the messenger mass scale, $M_\mathrm{mess} = kA_Z$, are given by
\begin{align}
\label{eq:GUTgaugino}
M_{a}     \simeq N_5\frac{\alpha_a}{4\pi} \frac{F_Z}{A_Z} \ ,
\end{align}
while the soft SUSY breaking masses squared of 
the MSSM scalar fields ($\phi$) are given by
\begin{align}
\label{eq:GUTsfermion}
m_{\phi}^2 \simeq  
\frac{2N_5}{16\pi^2}
\left(C_2(r_3^\phi)\alpha_3^2+C_2(r_2^\phi)\alpha_2^2 + \frac{3}{5}Q_Y^{\phi 2}\alpha_1^2\right)\,\frac{F_Z^2}{A_Z^2}\ ,
\end{align}
\cite{Dine:1993yw,Dine:1994vc,Dine:1995ag,Martin:1996zb}.
Here, we assume that there are $N_5$ pairs of the messenger multiplets. 
The upper limit on $N_5$ is about
$N_5+1 \lesssim 150/\log(M_\mathrm{GUT}/M_\mathrm{mess})$, which is imposed from the perturbativity of the coupling constants at the GUT scale, $M_\mathrm{GUT}$.
The index $a = 1,2,3$ corresponds to the MSSM gauge groups, U$(1)_Y$, SU$(2)_L$, and SU$(3)_c$, respectively. 
$\alpha_a$ are corresponding fine structure constants.
$C_2^\phi$ are the quadratic Casimir invariants of the representations $r_{a}^\phi$, 
and $Q_Y^\phi$ is the U$(1)_Y$ charges of the scalar field $\phi$.
We have assumed $F_Z/kA_Z^2 \ll 1$.
The SUSY breaking trilinear $A$-terms vanish at the messenger scale.
The mediated SUSY breaking masses are independent of the coupling constant $k$ at the leading order.%
\footnote{The $k$ dependence of the soft masses appears in higher order terms, $\order{F_Z^2/k^2A_Z^4}$, at the messenger scale.

The soft masses also has a logarithmic dependence on $k$ through the messenger mass, $M_{\mathrm{mess}} = k A_Z$.}
We call Eqs.\,\eqref{eq:GUTgaugino} and \eqref{eq:GUTsfermion} the GUT relation.

\begin{figure}[tbp]
	\centering
	\subcaptionbox{\label{fig:N5_1_mGMSB} $N_5 = 1$ }{\includegraphics[width=0.47\textwidth]{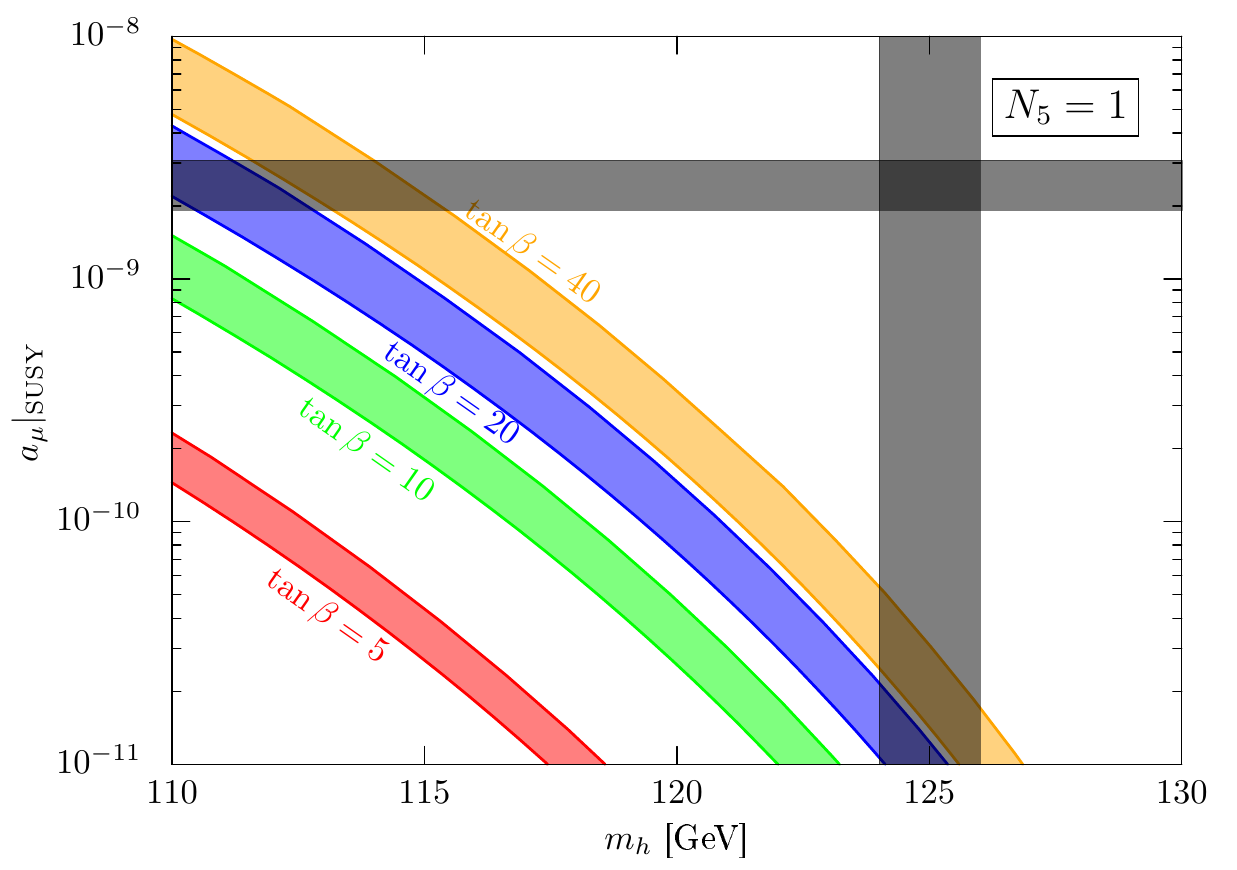}}
	\subcaptionbox{\label{fig:N5_5_mGMSB}$N_5 = 5$}{\includegraphics[width=0.47\textwidth]{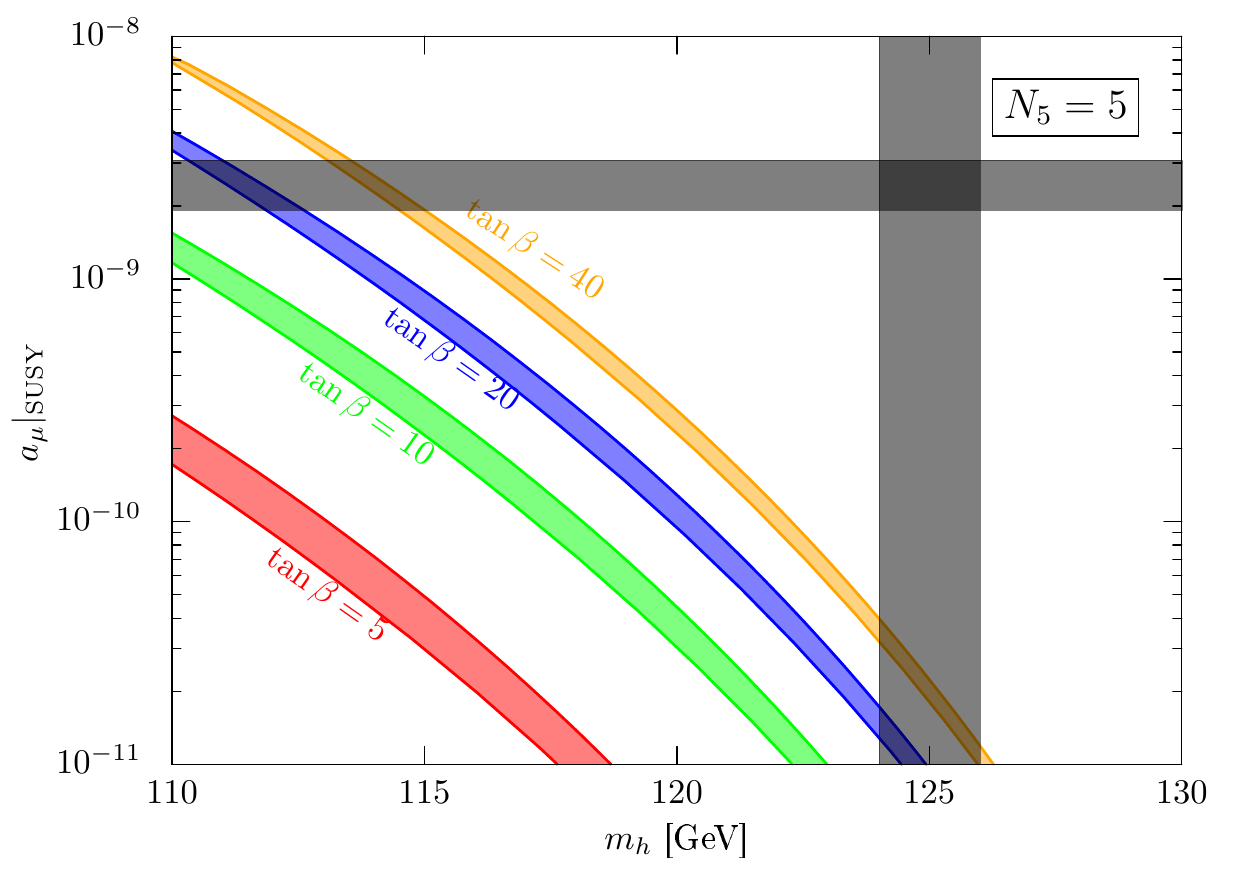}}
\caption{The predicted Higgs mass and $a_{\mu}$ in the minimal GMSB.
For each value of $\tan\beta$, we vary $F_Z/A_Z$ and $M_{\mathrm{ mess}} \in [10^4, 10^{16}]$\,GeV.
The vertical and the horizontal shaded regions correspond to the observed values of the Higgs boson mass and $a_\mu$, respectively.
For the Higgs mass constraint, we include the theoretical uncertainty of the prediction of the Higgs mass.
}
\label{fig:mGMSB}
\end{figure}

In Fig.\,\ref{fig:mGMSB},
we show the predicted Higgs boson mass and $a_\mu$ in the minimal GMSB for $N_5=1$ (left)
and $N_5 = 5$ (right).
In the figure, we vary $F_Z/A_Z$ and $M_{\mathrm{ mess}} \in [10^4, 10^{16}]$\,GeV for a given $\tan\beta$.
In our analysis, we have used the programs {\tt SOFTSUSY 4.1.10} \cite{Allanach:2001kg} to estimate the SUSY mass spectrum, {\tt FeynHiggs 2.18.0} \cite{Heinemeyer:1998yj,*Heinemeyer:1998np,*Degrassi:2002fi,*Frank:2006yh,*Hahn:2013ria,*Bahl:2016brp,*Bahl:2017aev,*Bahl:2018qog,*Bahl:2019hmm} for the Higgs mass calculation, and 
{\tt GM2Calc 1.7.5} \cite{Athron:2015rva} for the $a_\mu$ estimation.
In our analysis, we adopt the PDG average of the top mass measurement $m_t = 172.76 \pm 0.30$\,GeV \cite{Zyla:2020zbs}.
The Higgs mass is measured as $m_h = 124.97 \pm 0.24$\,GeV by the ATLAS  collaboration \cite{Aaboud:2018wps} and  $m_h = 125.38 \pm 0.16$\,GeV by the CMS collaboration \cite{Sirunyan:2020xwk}.
In addition to the experimental error of the Higgs mass measurement, there are theoretical uncertainties of the Higgs mass estimation originated from missing higher-order corrections and the experimental and the theoretical errors of the top mass~(see e.g., Ref.~\cite{Draper:2016pys}).
In this analysis, we assume the theoretical uncertainty of the Higgs mass boson is 1\,GeV.

As the figures show, $a_\mu|_\mathrm{SUSY}$
is below $10^{-10}$ when the Higgs boson mass is $m_h \simeq 125$\,GeV.
Thus, we find that the minimal GMSB fails to explain $a_\mu$ and the Higgs boson mass simultaneously.
Note that the ratio between the slepton masses and the squark masses deviates from the GUT relation for $F_Z/k A_Z^2 \to 1$. 
We have checked that the minimal GMSB model cannot explain $a_\mu$ and the Higgs boson mass simultaneously even in such a parameter region.

Before closing this subsection,
let us comment on 
the GMSB models with a messenger-Higgs mixing.
In the presence of the messenger-Higgs mixing,
a rather large trilinear $A$-term
can be generated~\cite{Chacko:2001km,Chacko:2002et,Evans:2011bea}. 
With a large $A$-term, 
the observed Higgs boson mass can be obtained 
for relatively light gluino/squarks~\cite{Evans:2011bea,Evans:2012hg,Kang:2012ra}.
In fact, $a_\mu$ and the observed Higgs boson mass can be explained simultaneously without violating the GUT relation~\cite{Evans:2012hg}.
The parameter region discussed in Ref.~\cite{Evans:2012hg} predicts a very light sparticles, 
which are severely constrained by the LHC searches.
In this paper, we focus on the case
without a messenger-Higgs mixing, and hence, the trilinear 
 $A$-terms vanish at the messenger scale as in the minimal setup of GMSB.

\subsection{GMSB without GUT relation}
As we have discussed, GMSB with the GUT relation does not explain 
the observed $a_\mu$ and the Higgs boson mass simultaneously.
In this section, we discuss the models with violation of the GUT relation.
The simplest model which violates the GUT relation is given by,
\begin{align}
\label{eq:GUTbreaking}
    W = (k_D Z' +M_D)\bar{\Phi}_D\Phi_{\bar{D}} 
    + (k_L Z'+M_L)\bar{\Phi}_{\bar{L}}\Phi_L
    \ ,
\end{align}
where the two types of the messenger multiplets have the independent mass parameters.
We also changed the spurious SUSY breaking
field to the one which only has the $F$-term VEV,
\begin{align}
    \langle Z'\rangle = F_{Z'}\theta^2\ .
\end{align}
In this model, the GUT relation is violated and the resultant GMSB soft masses at the messenger scale are modified to,
\begin{align}
\label{eq:gauginowoGUT}
&M_{3}   \simeq \frac{\alpha_3}{4\pi} 
\Lambda_{\mathrm{GMSB}}^D \ , \\
&M_{2}   \simeq \frac{\alpha_2}{4\pi} 
\Lambda_{\mathrm{GMSB}}^L \ , \\
&M_{1}   \simeq \frac{\alpha_a}{4\pi} \left(\frac{3}{5}
\Lambda_{\mathrm{GMSB}}^D
+ 
\frac{2}{5}
\Lambda_{\mathrm{GMSB}}^L
\right)\ , 
\end{align}
and
\begin{align}
\label{eq:sfermionwoGUT}
    m_{\phi}^2 \simeq \frac{2}{16\pi^2}
&\left(
C_2(r_3^\phi)\alpha_3^2
|\Lambda_{\mathrm{GMSB}}^{D}|^2
+C_2(r_2^\phi)\alpha_2^2
|\Lambda_{\mathrm{GMSB}}^{L}|^2 \right.
\nonumber\\
& \hspace{1cm}\left.
+ \frac{3}{5}Q_Y^{\phi2}\alpha_1^2
\left(
\frac{2}{5}
|\Lambda_{\mathrm{GMSB}}^{D}|^2
+
\frac{3}{5}
|\Lambda_{\mathrm{GMSB}}^{L}|^2
\right)
\right)\,\ .
\end{align}
Here, we have defined,
\begin{align}
\Lambda_{\mathrm{GMSB}}^{D} = \frac{k_DF_{Z'}}{M_D}
\ , \quad
\Lambda_{\mathrm{GMSB}}^{L} = \frac{k_LF_{Z'}}{M_L}\ .
\end{align}
Here, we have assumed $|k_{D,L}F_{Z'}/A_{Z_{D,L}}^2|\ll 1$.
With the violation of the GUT relation, it is possible to explain $a_\mu$ and the observed Higgs boson mass simultaneously by taking $|\Lambda_{\mathrm{GMSB}}^{D}| \gg |\Lambda_{\mathrm{GMSB}}^L|$ 
(see e.g., Ref.~\cite{Bhattacharyya:2018inr}).%
\footnote{For successful model,
the Higgs soft masses squared also require 
additional sources other than GMSB.
We will discuss this point in the next section.
}

The GUT violating messenger interactions, however, introduce new sources of CP violation. 
Unlike the messenger coupling in Eq\,\eqref{eq:messenger}, we can not eliminate all of the complex phases
of the parameters in Eq.\,\eqref{eq:GUTbreaking} 
by field redefinitions.
As a result, there is a relative phase
of $\order{1}$ between $\Lambda_{\mathrm{GMSB}}^D$ and $\Lambda_{\mathrm{GMSB}}^L$,
which propagates to the gaugino masses and $B$ through the renormalization group (RG) equations.
Once the gaugino masses and $B$ have relative phases of $\order{1}$, the resultant soft parameters can induce the non-vanishing electric dipole moments (EDMs).
In particular, the electron EDM, $d_e$, is roughly correlated with $a_\mu$
\begin{align}
\label{eq:EDM}
  \left|\frac{d_e}{e} \right|\sim  \frac{1}{2}\frac{m_e}{m_\mu^2}\times a_{\mu}|_\mathrm{SUSY}\sim 10^{-24}\,\mathrm{cm}\times\left(\frac{a_\mu|_\mathrm{SUSY}}{2\times 10^{-9}}\right)\  ,
\end{align}
where $e$ is the QED coupling constant and 
$m_e$ is the electron mass.%
\footnote{When either the Bino or the Wino decouples from the SUSY contributions to $a_\mu$, the gaugino mass contribution to the EDM can be suppressed if we can tune the complex phases of $\mu$ and $B$.} 
By comparing this equation with the current upper bound on the electron EDM given by ACME~\cite{Andreev:2018ayy},
\begin{align}
    \left|\frac{d_e}{e} \right|< 1.1 \times 10^{-29}\,{\mathrm{cm}}\ ,
\end{align}
we see that an accidental tuning is required.

\section{CP-Safe GMSB without GUT Relation}
\label{sec:Aligned}
In the above discussion, we have found that: 
\begin{itemize}
\item{GUT violating messenger coupling is required to explain the $a_\mu$ and the Higgs mass simultaneously}

\item{Naive GUT violation of the messenger coupling ends up with a too large electron EDM.}
\end{itemize}
In this section, we propose a model of the GUT violating messenger sector which avoids the CP violating phases.

\subsection{Alignment of CP phases}
\label{sec:CPsafeGUTbreaking}
To avoid the unwanted CP phases in the GUT violating messenger coupling, let us introduce two independent SUSY breaking fields, $Z_D$ 
and $Z_L$.
As we will see shortly,
they obtain the VEVs of 
\begin{align}
\label{eq:LDSUSYbreaking}
    &\langle Z_D \rangle  = A_{Z_D} + F_{Z_D} \theta^2\ ,\\
    & \langle Z_L \rangle  = A_{Z_L} + F_{Z_L} \theta^2 \ .
\end{align}
We will also see that it is possible to align all the phases of $A_{Z_{D,L}}$ and $F_{Z_{D,L}}$.
The Down-type messengers and the Lepton-type messengers couple to $Z_{D,L}$ via
\begin{align}
\label{eq:GUTbreaking2}
    W = k_D Z_D \bar{\Phi}_D\Phi_{\bar{D}}
+ k_L Z_L \bar{\Phi}_L\Phi_{\bar{L}} \ .
\end{align}
By appropriate phase redefinitions of $\bar{\Phi}_D\Phi_{\bar{D}}$ and 
$\bar{\Phi}_L\Phi_{\bar{L}}$, we can always take $k_{D,L}$ real-positive valued.
Thus, if we can prepare the SUSY breaking fields in Eq.\,\eqref{eq:LDSUSYbreaking} with all of $A_{Z_{D,L}}$ and $F_{Z_{D,L}}$ real-positive, we achieve the GUT violating messenger coupling without CP violation.

Now, let us discuss how we can prepare the SUSY breaking sector in which the phases of the VEVs of $Z_D$ 
and $Z_L$ are aligned. 
We assume that there are two independent SUSY breaking sectors where each SUSY breaking 
is caused by the $F$-component VEVs of $Z_{D}$ and $Z_{L}$, respectively.
We also assume that the mass scales of the two sectors are not very different.
Then, the effective theory
of the pseudo-flat directions $Z_{D,L}$ is described by the K\"ahler potential
and the superpotential,
\begin{align}
\label{eq:K}
    &K \simeq Z_D^\dagger Z_D - \frac{
    (Z_D^\dagger Z_D)^2 }{\Lambda_D^2}+ Z_L^\dagger Z_L - \frac{
    (Z_L^\dagger Z_L)^2 }{\Lambda_L^2}\ ,\\
    \label{eq:W}
    &W = w_D^2 Z_D + w_L^2 Z_L  + m_{3/2}M_{\mathrm{Pl}}^2\ .
\end{align}
Here, $m_{3/2}$ is the gravitino mass and $M_{\mathrm{Pl}}$ is the reduced Planck scale.
The mass parameters $w_{D,L}^2$
and $\Lambda_{D,L}^2 \in \mathbb{R}$ encapsulate the ultraviolet (UV) completion of the two SUSY breaking sectors.
We neglected the dimension $6$ or higher order terms.
Each SUSY breaking sector can be the low energy effective theory of, for example, the O'Raifeartaigh-type SUSY breaking model~\cite{ORaifeartaigh:1975nky}
(see also \cite{Kitano:2006wm,Kitano:2006wz,Ibe:2006rc,Ibe:2007km} and the Appendix~\ref{sec:HiggsSSS}).

We assume symmetries under the phase rotations of $Z_{D,L}$,
U$(1)_{D,L}$,
which are explicitly broken only by the mass parameters, $w_{D,L}^2$, respectively.
By giving U$(1)_{D,L}$ charges 
to $\bar{\Phi}_D\Phi_{\bar{D}}$
and $\bar{\Phi}_{\bar{L}}\Phi_{L}$, we forbid the mixings such as $Z_D\bar{\Phi}_{\bar{L}}\Phi_{L}$.
We also assume the $R$-symmetry which is broken only by $m_{3/2}$.
By U$(1)_{D,L}$ and the
$R$-symmetry rotation, 
we can take
\begin{align}
\label{eq:aligned}
    w_{D,L}^2 > 0\ , \qquad 
    m_{3/2}>0\ ,
\end{align}
without loss of generality. 

$A_{Z_{D,L}}$ are determined by the minimum of the scalar potential of $Z_{D,L}$,
\begin{align}
    V \simeq
     4 \frac{w_D^4}{\Lambda_D^2} |A_{Z_D}|^2
    + 4 \frac{w_L^4}{\Lambda_D^2} |A_{Z_L}|^2
    -2 (m_{3/2} w_D^2 A_{Z_D}+ h.c.)
    -2 (m_{3/2} w_L^2 A_{Z_L}+ h.c.)\ ,
\end{align}
where we neglected the dimension $8$ or higher order terms.
We also neglected the terms of $\mathcal{O}(m_{3/2}^2)$.
The above expansion is valid for $|A_{Z_{D,L}}|\ll \Lambda_{D,L}$.
The pseudo-flat directions have the positive mass terms with
\begin{align}
\label{eq:Zmass}
    m_{D,L}^2 \simeq 4 \frac{w_{D,L}^4}{\Lambda_{D,L}^2} > 0  \ , 
\end{align}
which is the generic feature of the O'Raifeartaigh models~\cite{Shih:2007av,Komargodski:2009jf,Evans:2011pz}.
With this scalar potential, we obtain the 
VEVs of the SUSY breaking field as \cite{Kitano:2006wm,Kitano:2006wz,Ibe:2006rc,Ibe:2007km},
\begin{align}
\label{eq:AVEV}
     A_{Z_{D,L}}\simeq \frac{m_{3/2} \Lambda_{D,L}^2}{2w_{D,L}^2} = \frac{\sqrt{3}\Lambda_{D,L}^2}{6M_\mathrm{Pl}}\ ,
\end{align}
and their $F$ components%
\footnote{In our convention, the superpotential contributes to the Lagrangian density as $\mathcal{L}_W = -\int d^2\theta W+ h.c$.
}
\begin{align}
F_{Z_{D,L}} \simeq w_{D,L}^{2}\ .
\end{align}
As we have aligned the 
phases of the parameters 
as in Eq.\,\eqref{eq:aligned}, this setup provides appropriate SUSY breaking fields in Eq.\,\eqref{eq:LDSUSYbreaking} with all the CP phases aligned, 
that is,
\begin{align}
    A_{Z_{D,L}}> 0 \ , \quad F_{Z_{D,L}} > 0 \ .
\end{align}

With the condition for the vanishing cosmological constant,
\begin{align}
    F_{Z_{D}}^2 + F_{Z_{L}}^2 - 3 m_{3/2}^2 M_{\mathrm{Pl}}^2 \simeq 0\ ,
\end{align}
we parametrize as
\begin{align}
&F_{Z_{D,L}} \simeq \sqrt{3} \kappa_{D,L} m_{3/2} M_{\mathrm{Pl}}\ , \\
&\kappa_{D,L} \equiv \frac{F_{Z_{D,L}}}{\sqrt{F_{Z_{D}}^2+ F_{Z_{L}}^2}}\ .
\end{align}
Accordingly, the combinations relevant for the GMSB spectrum
in Eqs.\,\eqref{eq:gauginowoGUT}-\eqref{eq:sfermionwoGUT} are written as,
\begin{align}
    \Lambda^{D,L}_{\mathrm{GMSB}} \simeq 6 m_{3/2} \times\frac{\kappa_{D,L}M_{\mathrm{Pl}}^2}{\Lambda_{D,L}^2}\ .
\end{align}
For successful GMSB, we require $\Lambda_{\mathrm{GMSB}}^{D,L} = 10^{5\mbox{--}6}$\,GeV, and hence,
\begin{align}
\label{eq:LambdaDL}
    \Lambda_{D,L} \simeq 6\times 10^{15}\,{\mathrm{GeV}}\times\left(
    \frac{m_{3/2}}{\mathrm{GeV}}\right)^{1/2}\left(\frac{\kappa_{D,L}\times 10^6\,\mathrm{GeV}}{\Lambda_{\mathrm{GMSB}}^{D,L}}\right)^{1/2}\ .
\end{align}
To explain $a_\mu$ and the Higgs boson mass simultaneously, we take $\Lambda_{\mathrm{GMSB}}^D/\Lambda_{\mathrm{GMSB}}^L \simeq 5$--$\mathrm{10}$, which is achieved for, for example, $\kappa_D \simeq 1$ and $\kappa_L \simeq 0.1$--$0.2$ if $\Lambda_D \sim \Lambda_L$.

Finally, let us discuss  the fermion components of $Z_{D,L}$.
As $Z_{D,L}$ break global SUSY independently, the fermion components are massless goldstini, $\tilde{G}_{D,L}$.
One linear combination of them becomes the gravitino with a mass, $m_{3/2}$,
and the other obtains a mass, $2m_{3/2}$, through the super-Higgs mechanism~\cite{Cheung:2010mc}.

\subsection{Vacuum Stability}
Since the SUSY breaking fields couple to the messenger fields 
in superpotential Eq.\,\eqref{eq:GUTbreaking2}, there is a supersymmetric vacuum at
\begin{align}
\label{eq:SUSYVEV}
    \langle{\bar{\Phi}_D\Phi_D}\rangle = - 
    w_D^2/k_D\ , \qquad
    \langle{\bar{\Phi}_{\bar{L}}\Phi_L}\rangle = - 
    w_L^2/k_L\ , 
\end{align}
with vanishing $\langle Z_{D,L}\rangle$.
Hence, the SUSY breaking vacuum in Eq.\,\eqref{eq:AVEV} is at best meta-stable \cite{Kitano:2006wz}.
The squared masses of the messenger scalars 
around the meta-stable vacuum is given by,
\begin{align}
    \mathcal{M}_{D,L}^2=\left(\begin{array}{cc}
         k_{D,L}^2 A_{Z_{D,L}}^2 & k_{D,L} F_{Z_{D,L}} \\
         k_{D,L} F_{Z_{D,L}} &  k_{D,L}^2 A_{Z_{D,L}}^2
    \end{array}
    \right) \ .
\end{align}
Hence, the meta-stability condition, $\det\mathcal{M}_{D,L}^2 > 0 $, leads to
\begin{align}
     k_{D,L} >\frac{F_{Z_{D,L}}}{A_{Z_{D,L}}^2} \simeq 3\times 10^{-8}\times \kappa_{D,L}
    \left(\frac{m_{3/2}}{1\,\mathrm{GeV}}\right)\left(\frac{10^{16}\,\mathrm{GeV}}{\Lambda_{D,L}}\right)^4\ .
\end{align}

The couplings to the messenger fields also induce the Coleman-Weinberg potentials to the pseudo flat directions,
\begin{align}
{\mit\Delta}V(A_{Z_{D,L}}) = d_{D,L}w_{D,L}^4\times \frac{k_{D,L}^2}{16\pi^2} \log\frac{|A_{Z_{D,L}}|^2}{\Lambda_{D,L}^2} 
\ ,
\end{align}
where $d_D = 3$ and $d_L=2$.
These terms contribute to the mass matrix of $A_{Z_{D,L}}$ in the $(A_{Z_{D,L}}, A_{Z_{D,L}}^\dagger)$ basis,
\begin{align}
  \mathcal{M}_{Z_{D,L}}^2=  \frac{4w_{D,L}^4}{\Lambda_{D,Z}^2}
    \left(
    \begin{array}{cc}
        1 & - \displaystyle{\frac{d_{D,L}k_{D,L}^2}{64\pi^2}\frac{\Lambda_{D,L}^2}{Z_{D,L}^2}} \\ 
        - \displaystyle{\frac{d_{D,L}k_{D,L}^2}{64\pi^2}\frac{\Lambda_{D,L}^2}{Z_{D,L}^2}} & 1
    \end{array}
    \right)\ .
\end{align}
As a result, another meta-stability condition, $\det\mathcal{M}_{Z_{D,L}}^2 > 0$, leads to
\begin{align}
k_{D,L} < \frac{4\pi \Lambda_{D,L}}{\sqrt{3 d_{D,L}} M_\mathrm{Pl}}
    \simeq 10^{-2} \times  \left(\frac{\Lambda_{D,L}}{10^{16}\,\mathrm{GeV}}\right)
    \ .
\end{align}

In summary, the meta-stable vacuum conditions restrict the range of the messenger scale, $M_{\mathrm{mess}}=k_{D,L} A_{Z_{D,L}} $, in,
\begin{align}
6 m_{3/2} \times\frac{\kappa_{D,L}M_{\mathrm{Pl}}^2}{\Lambda_{D,L}^2}< 
M_{\mathrm{mess}} < \frac{2\pi \Lambda_{D,L}^3}{3d_{D,L}^{1/2}M_{\mathrm{Pl}}^2}\ .
\end{align}
Hence, the messenger scale lies in the range,
\begin{align}
\label{eq:Mmess}
3\times 10^5\,\mathrm{GeV}\times \left(\frac{m_{3/2}}{\mathrm{GeV}}\right)
\left(\frac{10^{16}\,\mathrm{GeV}}{\Lambda_{D,L}}\right)^2
\lesssim
M_{\mathrm{mess}} \lesssim
3\times 10^{11}\,\mathrm{GeV}\times \left(\frac{\Lambda_{L,D}}{10^{16}\,\mathrm{GeV}}\right)^3 \ .
\end{align}

Finally, let us discuss the tunneling rate of the meta-stable vacuum into the supersymmetric vacuum in Eq.\,\eqref{eq:SUSYVEV}.
The VEVs in Eq.\,\eqref{eq:AVEV} are much larger than the VEVs of the messenger fields in Eq.\,\eqref{eq:SUSYVEV}.
The displacement between the 
meta-stable vacuum and the unwanted color-breaking supersymmetric vacuum is of 
order of ${\mit \Delta}A_{D,L} \sim {\Lambda_{D,L}^2/M_\mathrm{Pl}}$.
The tunneling rate per unit volume, $\Gamma/V \propto e^{-S_E}$, is estimated in Ref.\,\cite{Hisano:2008sy} 
where 
\begin{align}
    S_E \sim 8\pi^2
    \left(\frac{{\mit \Delta}A_{D,L}}{w_{D,L}}\right)^4 \sim 8\pi^2\times
    \left(\frac{\Lambda_{D,L}}{\Lambda_{\mathrm{GMSB}}^{D,L}}\right)^2\ .
\end{align}
Therefore, the meta-stable vacuum is stable enough, for example, for $\Lambda_{D,L}\gg \Lambda_\mathrm{GMSB}^{D,L}$ so that $S_E \gtrsim 500$.%
\footnote{In our model, we assume that the messenger fields are heavy and not thermally produced after inflation. In such a case, the pseudo-flat directions are neither thermalized.}

\subsection{GUT violating messenger multiplets}
In Eq.\,\eqref{eq:GUTbreaking2},
we assume the Down-type and the Lepton-type messenger multiplets which couple to $Z_D$ and $Z_L$, respectively.
The simplest realization of such GUT violating messenger multiplets is
to consider the 
product group unification models~\cite{Izawa:1997he,Yanagida:1994vq,Hisano:1995hc,Hotta:1996qb,Hotta:1996pn}.
The product group unification is motivated to solve the infamous doublet-triplet splitting problem of the Higgs multiplets in the conventional GUT.

As a concrete example, let us consider the product group GUT model based on the GUT gauge group, SU$(5)\times$U$(2)_{H}$~\cite{Izawa:1997he,Ibe:2003ys}.%
\footnote{See Ref.\,\cite{Evans:2020fmh} for the status of the proton lifetime in this model. The predicted proton lifetime also depends on the origin of the leptons in the product group unification~\cite{Ibe:2019ifm}.}
In this model, 
SU$(5)\times$U$(2)_{H}$ is spontaneously broken down to the MSSM gauge groups by the VEV of the
chiral multiplets of the vector-like bi-fundamental representation, $(Q,\bar{Q})$. Their VEVs are,
\begin{align}
    \langle Q \rangle = 
    \left(\begin{array}{cc}
         v &0  \\
         0 & v \\
         0 & 0 \\
         0 & 0 \\
         0 & 0
    \end{array}\right)\ , 
    \qquad 
     \langle \bar{Q} \rangle = 
    \left(\begin{array}{ccccc}
         v &0 & 0 & 0 & 0\\
         0 & v & 0 & 0 & 0    \end{array}\right)\ , 
\end{align}
where $v$ denotes the VEV of the order of the GUT scale.
In this model, the nominal coupling unification of the MSSM gauge group at the GUT scale is explained in the strong coupling limit of U(2)$_{H}$ gauge interaction.

Now, let us introduce a messenger multiplet $(\Phi_5,\bar{\Phi}_5)$
in the (anti-)fundamental representation of SU$(5)$, a vector-like multiplet $(\Phi_2,\bar{\Phi}_2)$ in the (anti-)fundamental representation of SU$(2)_{H}$.
The messenger fields couple to the SUSY breaking field $Z_D$ through,
\begin{align}
W_D = k_D Z_D \bar{\Phi}_5\Phi_5 + \lambda\bar{\Phi}_5 Q \Phi_2 +
\bar{\lambda}\bar{\Phi}_2 \bar{Q} \Phi_5 \ ,
\end{align}
where $\lambda$ and $\bar{\lambda}$ are coupling constants.
Here, we can take all the parameters real-positive by field redefinitions.
Due to the second and third terms, the doublet components in $(\Phi_5, \bar{\Phi}_5)$ obtain masses of the GUT scale, and decouple.
Thus, $W_D$ provides the Down-type messenger in Eq.\,\eqref{eq:GUTbreaking2}.

The Lepton-type messengers can be also obtained by introducing the SU(2)$_{H}$ doublet 
$(\Phi_L, \bar{\Phi}_{\bar{L}})$ with U$(1)_{H}$ changes, $\mp 1/2$.
By assuming that
$(\Phi_L, \bar{\Phi}_{\bar{L}})$ couple to $Z_L$,
\begin{align}
    W_L = k_L Z_L \bar{\Phi}_{\bar{L}}\Phi_L \ ,
\end{align}
this sector results in the Lepton-type messenger in Eq.\,\eqref{eq:GUTbreaking2}
since  SU$(2)_{L}$ and U$(1)_{Y}$
of the SM corresponds to 
the diagonal subgroups of the SU$(5)$ and SU$(2)_H$ and U$(1)_H$.
In this way, we obtain the effective GUT violating messenger multiplets in Eq.\,\eqref{eq:GUTbreaking2}.

\section{Sweet Spot Supersymmetry}
\label{sec:SSS}
In the previous section,
we show how to achieve the gaugino masses whose CP phases 
are aligned while the GUT relation is violated.
To discuss the SUSY CP problem,
however, we also need to specify the origin of the $\mu$-term
as well as the $B$-term.

Also note that we are interested in the model with light sleptons and heavy squarks to 
explain  $a_\mu$ and the Higgs boson mass simultaneously.
In this case, the large squark masses induce the large Higgs soft masses squared, $m_{H_u}^2$ and $m_{H_d}^2$, at the TeV scale through the RG running.
With large $m_{H_{u,d}}^2$, the required size of the $\mu$-term is also large to achieve the correct electroweak symmetry breaking vacuum.
With a large $|\mu|$ term, the stau tends to be light and causes the stability problem~\cite{Hisano:2010re}.

To avoid a too large $\mu$-term, we introduce additional contributions to $m_{H_{u,d}}^2$ in addition to the GMSB contributions~\cite{Kitano:2006wm,Kitano:2006wz,Ibe:2006rc,Ibe:2007km} 
(see also Refs.~\cite{Ibe:2012qu,Yanagida:2017dao,Bhattacharyya:2018inr}).
The additional contributions offset the RG contributions.
In summary, 
for successful explanation of $a_\mu$ and the Higgs boson mass, we consider models with:
\begin{itemize}
    \item The mechanism which generates $\mu$-term without causing new CP phase
    \item The additional source of $m_{H_{u,d}}^2$ other than GMSB to achieve a small $\mu$-term.
\end{itemize}

\subsection{Higgs mass parameters}
As developed in Refs.~\cite{Kitano:2006wm,Kitano:2006wz,Ibe:2006rc,Ibe:2007km},
we can generate the $\mu$-term and the additional Higgs soft masses squared simultaneously by coupling the Higgs doublets with a SUSY breaking sector.
Here, we assume that the Higgs doublets couple to the SUSY breaking sector of $Z_D$.
Then, the direct coupling induces
the effective K\"ahler potential,
\begin{align}
\label{eq:Kmu}
    K= \frac{Z_{D}^\dagger}{\Lambda_\mu} H_u H_d + h.c. - \frac{Z_{D}^\dagger Z_{D}}{\Lambda_u^2} H_u^\dagger H_u  
     - \frac{Z_{D}^\dagger Z_{D}}{\Lambda_d^2} H_d^\dagger H_d\ .
\end{align}
The cutoff parameters
$\Lambda_{u,d}^2$ are real valued by definition, 
while we can take $\Lambda_\mu$ real-positive by the Peccei-Quinn (PQ) rotation of $H_uH_d$.
We discuss the details of the origin of the K\"ahler potential
in the Appendix.~\ref{sec:HiggsSSS},
where we find $\Lambda_{u,d}^2 > 0 $.
The resultant $\mu$-term and the additional Higgs soft masses squared are given by,%
\footnote{We define the phase of the $\mu$-term to be $\mathcal{L} = \mu \int d\theta^2 H_u H_d + h.c$.}
\begin{align}
\label{eq:mu}
    & \mu \simeq \frac{F_{Z_D}}{\Lambda_\mu}
    \ , \\
    & \delta m_{H_{u,d}}^2 
    \simeq\frac{F_{Z_D}^2}{\Lambda_{u,d}^2}\ ,
\end{align}
where $\mu$ is real-positive.
For successful explanation of $a_\mu$, we require that  $\mu$ is within a TeV range, and hence,
\begin{align}
\label{eq:Lambdamu}
\Lambda_\mu = 8 \times 10^{15}\,\mathrm{GeV}\times  \kappa_{D}
\left(\frac{m_{3/2}}{\mathrm{GeV}}\right)
\left(\frac{500\,\mathrm{GeV}}{\mu}\right)\ .
\end{align}
Similarly, the requirement that $\delta m_{H_{u,d}}^2$ is in a few TeV range leads to
\begin{align}
\label{eq:Lambdaud}
\Lambda_{u,d} \simeq 10^{15}\,\mathrm{GeV}\times \kappa_{D}
\left(\frac{m_{3/2}}{\mathrm{GeV}}\right)
\left(\frac{3\,\mathrm{TeV}}{\delta m_{H_{u,d}}}\right)\ .
\end{align}
With 
the closeness of $\Lambda_{L,D,\mu,u,d}$ 
for $m_{3/2} = \order{1}$\,GeV,
we call this scenario the (extended) Sweet Spot Supersymmetry which is originally proposed in Ref.~\cite{Ibe:2007km}.

In Eq.\,\eqref{eq:Kmu}, we assumed that 
the Higgs doublets do not couple to $Z_L$.
Such a model is possible by combining the $Z_D$ phase rotation symmetry in Eq.\,\eqref{eq:K} with the PQ phase rotation (see also the Appendix~\ref{sec:HiggsSSS}).
The same symmetry also forbids the terms such as $Z_{D}^{(\dagger)} H_{u,d}^\dagger H_{u,d}$ and $Z_D^\dagger Z_D H_uH_d$.
As a result, the $B$-term from the K\"ahler potential in Eq.\,\eqref{eq:Kmu} is,%
\footnote{The $B$-term is defined by $\mathcal{L} = - B  \mu H_u H_d + h.c.$}
\begin{align}
\label{eq:Bmu}
    B \mu = \frac{F_{Z_D}}{\Lambda_{\mu}} \times \left(\frac{ A_{Z_D} F_{Z_D}}{\Lambda_u^2}+\frac{ A_{Z_D} F_{Z_D}}{\Lambda_d^2}
    \right) = 
\frac{F_{Z_D}}{\Lambda_{\mu}} \times \left(\frac{\sqrt{3}\Lambda_{D}^2 }{6\Lambda_u^2}+
\frac{\sqrt{3}\Lambda_{D}^2 }{6\Lambda_d^2}
    \right)\frac{F_{Z_D}}{M_{\mathrm{Pl}}}\ll \mu^2\ .
\end{align}
Note that these contributions do not bring CP violating phases.
Similarly, the induced $A$-terms are also suppressed and do not have CP violating phases.
Since these $A$ and $B$-terms are harmless and overwhelmed by the RG contributions at the electroweak scale,
we neglect them in the following analysis.

In the Appendix~\ref{sec:HiggsSSS}, we discuss a perturbative UV completion of the effective K\"ahler potentials in Eqs.\,\eqref{eq:K} and \eqref{eq:Kmu}.
When the Higgs doublets couple to $Z_D$,%
\footnote{It is also possible that the Higgs doublets couple to $Z_L$ instead of $Z_D$.}
the cutoff scales are given by Eqs.\,\eqref{eq:ZZZZ}, \eqref{eq:ZZHH}, and \eqref{eq:ZHH};
\begin{align}
\label{eq:cutoff}
    \frac{1}{\Lambda_{D}} = \frac{\lambda^2}{2\sqrt{3}(4\pi)}\frac{1}{M_*} \ ,  \qquad
         \frac{1}{\Lambda_u} \simeq \frac{\lambda h}{4\pi}\frac{1}{M_{*}}\tilde{g}^{1/2}
    \ ,  \qquad
         \frac{1}{\Lambda_d} \simeq \frac{\lambda \bar{h}}{4\pi}\frac{1}{M_{*}} \tilde{g}^{1/2}
    \ ,  \qquad 
    \frac{1}{\Lambda_\mu} = \frac{\lambda h \bar{h}}{(4\pi)^2 M_{*}} \tilde{f}\ .
\end{align}
Here, the coupling constants, $\lambda$, $h$ and $\bar{h}$
defined in the Appendix~\ref{sec:HiggsSSS} are 
taken real-positive without loss of generality.
The coefficient functions $\tilde{f}$ and $\tilde{g}$ are given in Eqs.\,\eqref{eq:f} and \eqref{eq:g} (see also Fig.\,\ref{fig:functions}). 
Note that the ratio, $\tilde{g}^{1/2}/|\tilde{f}|$, is of $\order{1}$ for a wide range of parameters.
The mass parameter $M_*$ denotes the scale at which the higher dimensional operators in Eqs.\,\eqref{eq:K} and \eqref{eq:Kmu} are generated.
The result shows that the generated $\mu$-parameter is parametrically smaller
than $\delta m_{H_{u,d}}$ by an order of magnitude when $h$ and $\bar{h}$ are of $\order{1}$.
This small hierarchy between the $\mu$-parameter
and  $\delta m_{H_{u,d}}$ 
is desirable 
for the simultaneous explanation of $a_\mu$ and the Higgs boson mass.

\begin{figure}[t]
\centering
	\includegraphics[width=0.5\hsize,clip]{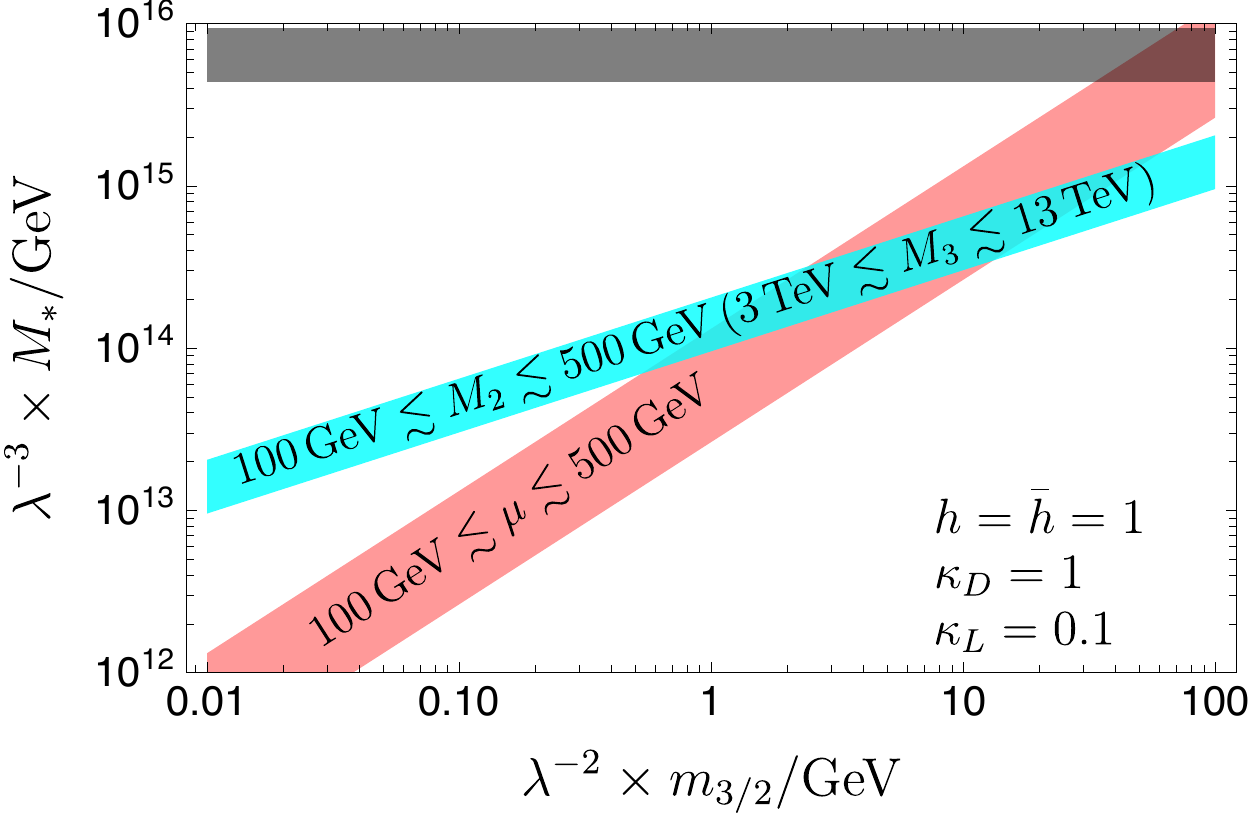}
	\caption{The parameter region satisfying Eq.\,\eqref{eq:Mstar1} (cyan) and Eq.\,\eqref{eq:Mstar2} (pink) for given range of $M_2$ and $\mu$. 
	We take $h = \bar{h}=1$
to avoid too large hierarchy between the $\mu$-parameter and $\delta m_{H_{u,d}}$.
We also fix $\kappa_D =1$ and $\kappa_L=0.1$, which is motivated to explain $a_\mu$ and the Higgs boson mass simultaneously.
In the gray shaded region, 
the VEV of $A_{Z_D}$ becomes too large (see Eq.\,\eqref{eq:lambda}).}
	\label{fig:SS}
\end{figure}

By combining Eq.\,\eqref{eq:cutoff} with Eq.\,\eqref{eq:LambdaDL}
and \eqref{eq:Lambdamu}, 
we find that the mediation scale is,
\begin{align}
\label{eq:Mstar1}
    &M_* \simeq 10^{14}\,\mathrm{GeV}\times \lambda^2
    \left(\frac{m_{3/2}}{1\,\mathrm{GeV}}\right)^{1/2}
    \left(\frac{\kappa_D\times 10^6\,\mathrm{GeV}}{\Lambda_{\mathrm{GMSB}}^D}\right)^{1/2}\ , \\
    \label{eq:Mstar2}
    &M_*\simeq 5\times 10^{13}\,\mathrm{GeV}\times \kappa_D\lambda h\bar{h}\tilde{f}
    \left(\frac{m_{3/2}}{1\,\mathrm{GeV}}\right)
    \left(\frac{ 500\,\mathrm{GeV}}{\mu}\right)\ .
\end{align}
In Fig.\,\ref{fig:SS}, we show the parameter region satisfying these conditions
for $\kappa_D = 1$, $\kappa_L=0.1$ and $h=\bar{h}=1$.
We also take the argument of $\tilde{f}(x)$ to be $1$.
The shaded bands correspond to,
 $100\,\mathrm{GeV}\lesssim M_2\lesssim 500$\,GeV
($3\,\mathrm{TeV}\lesssim M_3 \lesssim 13$\,TeV),
and  $100\,\mathrm{GeV}\lesssim \mu\lesssim 500$\,GeV, respectively.
In each band, the upper boundary corresponds to the lower values of $M_2$ or $\mu$.
The two region overlaps when the gravitino mass and the mediation scale $M_*$ satisfy
\begin{align}
    &m_{3/2} \simeq 7\,\mathrm{GeV}\times \frac{\lambda^2}{\kappa_Dh^2\bar{h}^2\tilde{f^2}}\left(\frac{\mu}{500\,\mathrm{GeV}}\right)^2
    \left(\frac{10^6\,\mathrm{GeV}}{\Lambda_{\mathrm{GMSB}}^D}\right)\ , 
    \label{eq:sweetspotm32}
    \\
    &M_* \simeq 2\times 10^{14}\,\mathrm{GeV}\times
    \frac{\lambda^3}{h \bar{h}\tilde{f}}\left(\frac{\mu}{500\,\mathrm{GeV}}\right)
    \left(
    \frac{10^6\,\mathrm{GeV}}{\Lambda_{\mathrm{GMSB}}^D}
    \right)\ .
    \label{eq:sweetspotMstar}
\end{align}
Therefore, we find that sweet spot 
is at around
$m_{3/2}= \lambda^2\times\order{1}$\,GeV and $M_*=\lambda^3\times\order{10^{14}}$\,GeV for $h$, $\bar{h}$ of $\order{1}$.

\subsection{Renormalization Group Analysis}
In the present model, 
the additional Higgs soft-mass squared and the $\mu$-term
are generated at the scale $M_*$, which is independent from $M_\mathrm{mess}$.
From Eqs.\,\eqref{eq:sweetspotMstar} and \eqref{eq:sweetspotm32},
we find 
\begin{align}
    M_* \simeq 2 \times 10^{13}\,\mathrm{GeV}\times h^2 \bar{h}^2 \kappa_D^{3/2}\tilde{f}^2
    \left(\frac{m_{3/2}}{1\,\mathrm{GeV}}\right)^{3/2}
    \left(\frac{500\,\mathrm{GeV}}{\mu}\right)^2 \left(\frac{\Lambda_\mathrm{GMSB}^D}{10^6\,\mathrm{GeV}}\right)^{1/2}\ .
\end{align}
Since we are mostly interested in the case $m_{3/2} = \order{100}$\,MeV to $\order{1}$\,GeV, we take 
$10^{11}\,\mathrm{GeV}\lesssim M_* \lesssim 10^{14}$\,GeV in the following analysis.
The gaugino masses and the sfermion masses are, on the other hand, generated at the messenger scale,
$M_{\mathrm{mess}}$,
which is assumed to be common 
for the Down-type and the Lepton-type messengers for simplicity.
The two step mediation at $M_*$ and $M_\mathrm{mess}$
predicts a peculiar spectrum~\cite{Ibe:2007km}.

\begin{figure}[tbp]
	\centering
	\subcaptionbox{\label{fig:RGE} RG evolution }{\includegraphics[width=0.52\textwidth]{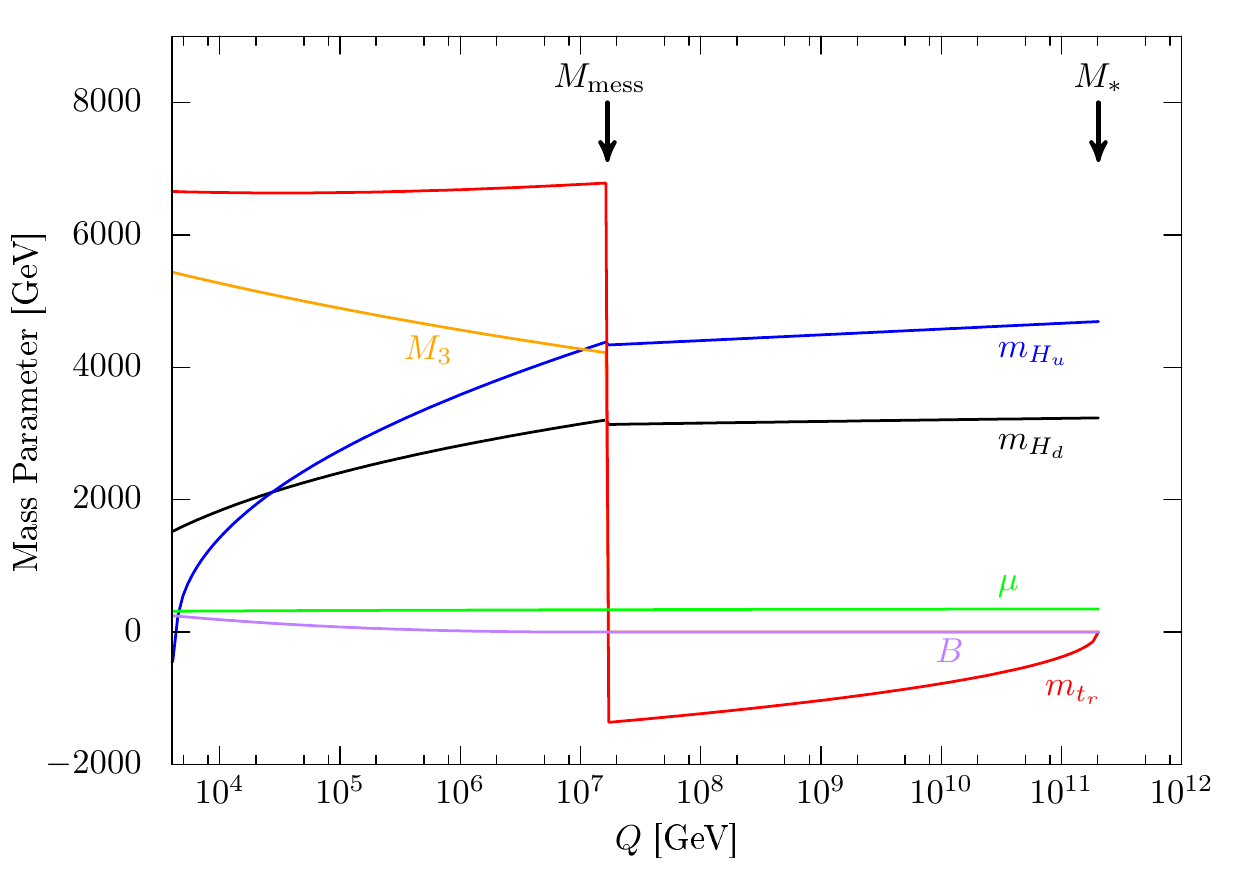}}
	\subcaptionbox{\label{fig:spectrum}$N_5 = 5$}{\includegraphics[width=0.44\textwidth]{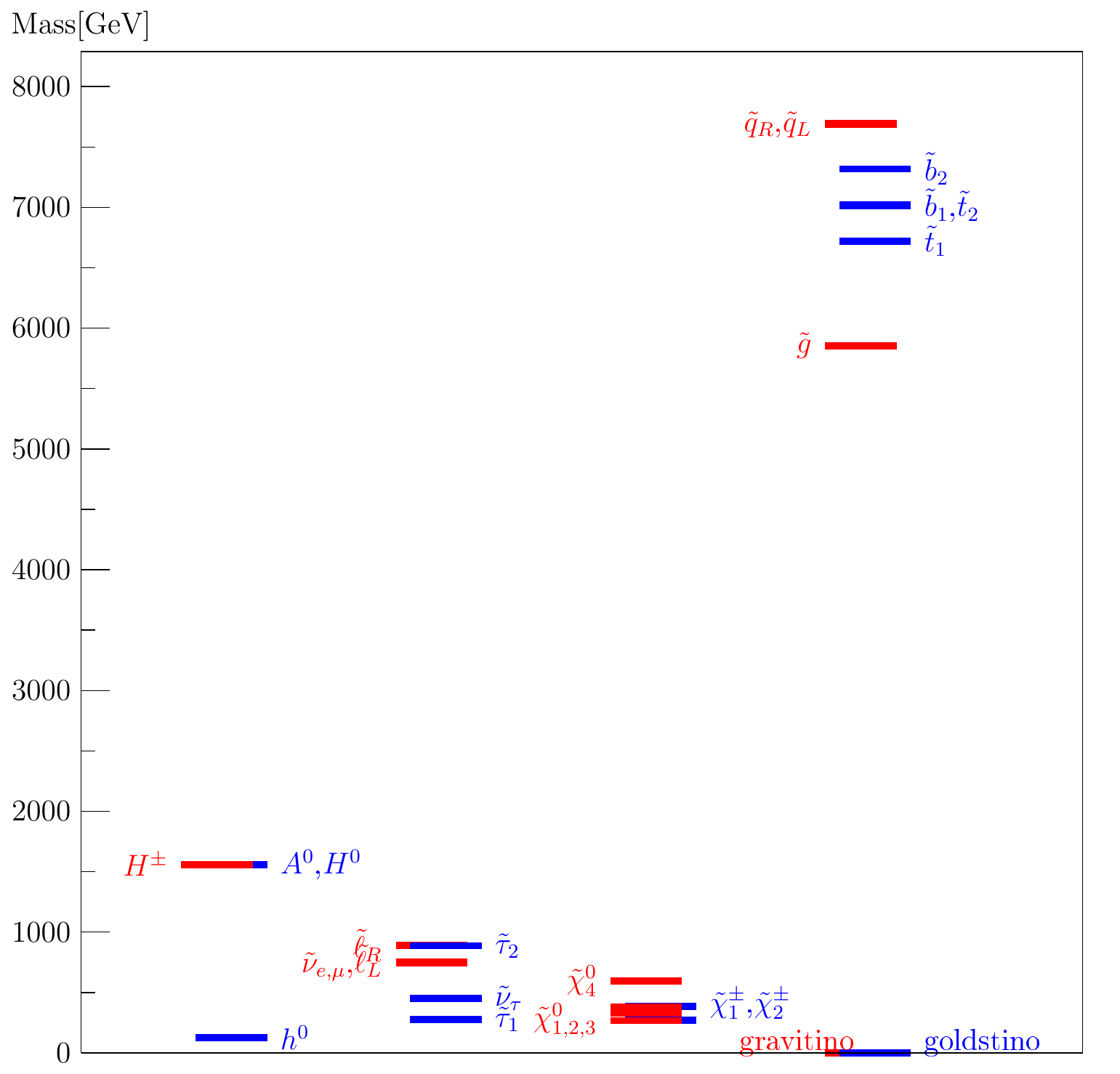}}
\caption{(a): The RG evolution of the SUSY parameters and (b): the mass spectrum of the sample point.
We take a sample point of $\Lambda^{L}_{\rm GMSB} = 124$ TeV, 
$\Lambda^{D}_{\rm GMSB} = 894$ TeV,
$M_{\rm mess} = 1.67 \times 10^7$ GeV,
$\Delta m_{H_u}^2 = 2.2\times 10^{7}$ GeV${}^2$,
$\Delta m_{H_d}^2 = 1.05\times 10^{7}$ GeV${}^2$ and
$M_{*} = 2 \times 10^{11}$ GeV, which predicts $\tan\beta = 41$, $m_{h^0} = 124.8$ GeV and $a_{\mu}|_{\rm SUSY} = 2.2 \times 10^{-9}$.
}
\label{fig:example}
\end{figure}

In Fig.~\ref{fig:RGE}, we show the RG running of the soft parameters.
As the figure shows, 
$\delta m_{H_{u,d}}^2$ generated at $M_*$ offset the negative RG contributions to $m_{H_{u,d}}^2$ at the TeV scale.
This feature makes a small $\mu$-parameter compatible with successful electroweak symmetry breaking.
It also shows that 
$\delta m_{H_{u,d}}^2$ give negative contributions to the soft sfermion masses of the third generation.
We also show an example of the spectrum which explains the $a_\mu$ and the observed Higgs boson mass simultaneously.%
\footnote{We provide the mass spectrum calculator and some sample spectra at
\url{https://member.ipmu.jp/satoshi.shirai/sweetspot/sweetspotSUSY.php}.
}

As we have seen, $B \mu$ is dominated by RG contributions from 
the gaugino mass through the RG evolution, and hence, $B \mu$ is not a free parameter.
Besides, there is no large contributions to $B \mu$ from  subdominant gravity mediation effects.
Thus, we take $B \mu$ to be vanishing above the messenger scale.
Accordingly, $\tan\beta$ is not a free parameter but is a prediction.

\begin{figure}[tbp]
	\centering
	\subcaptionbox{\label{fig:all_pdf} Without collider constraint. }{\includegraphics[width=0.47\textwidth]{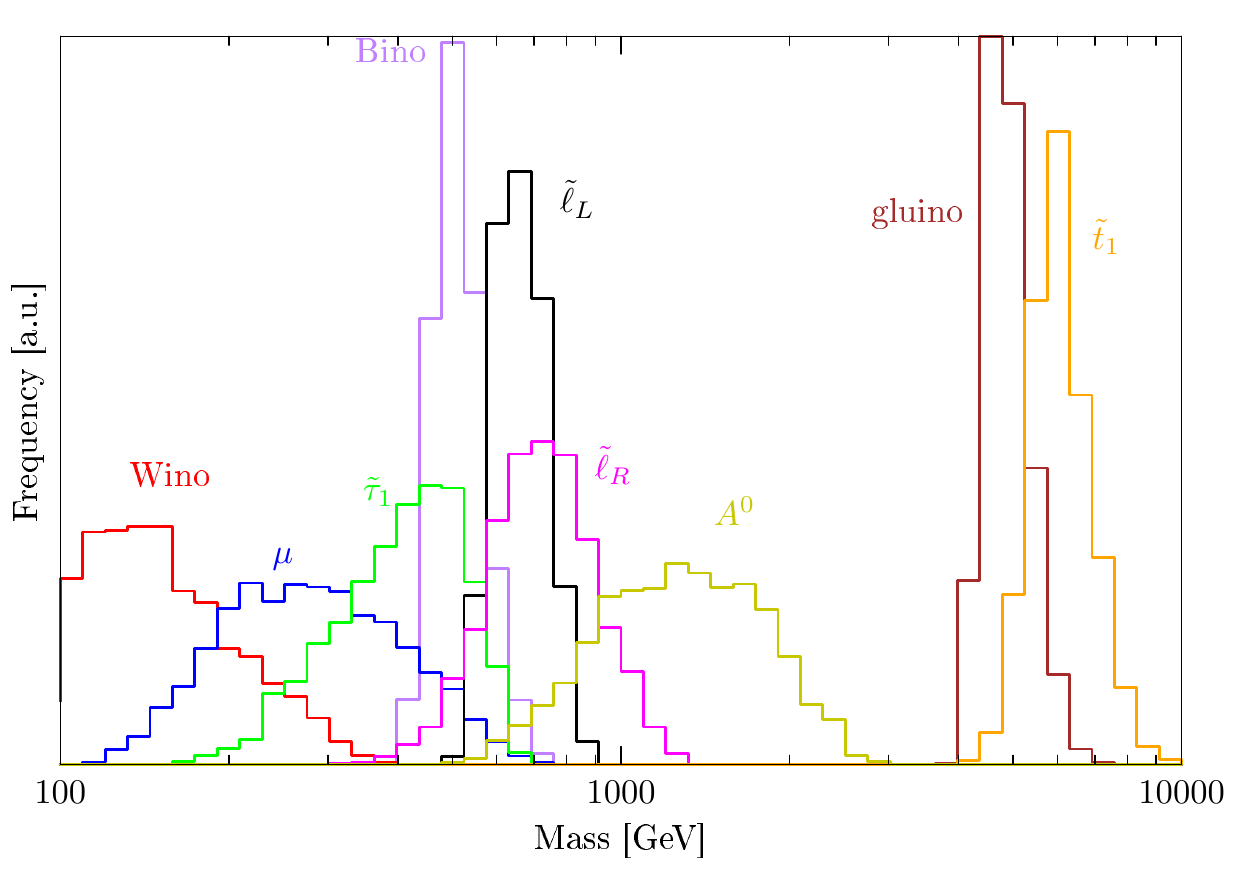}}
	\subcaptionbox{\label{fig:collider_pdf} With collider constraint.}{\includegraphics[width=0.47\textwidth]{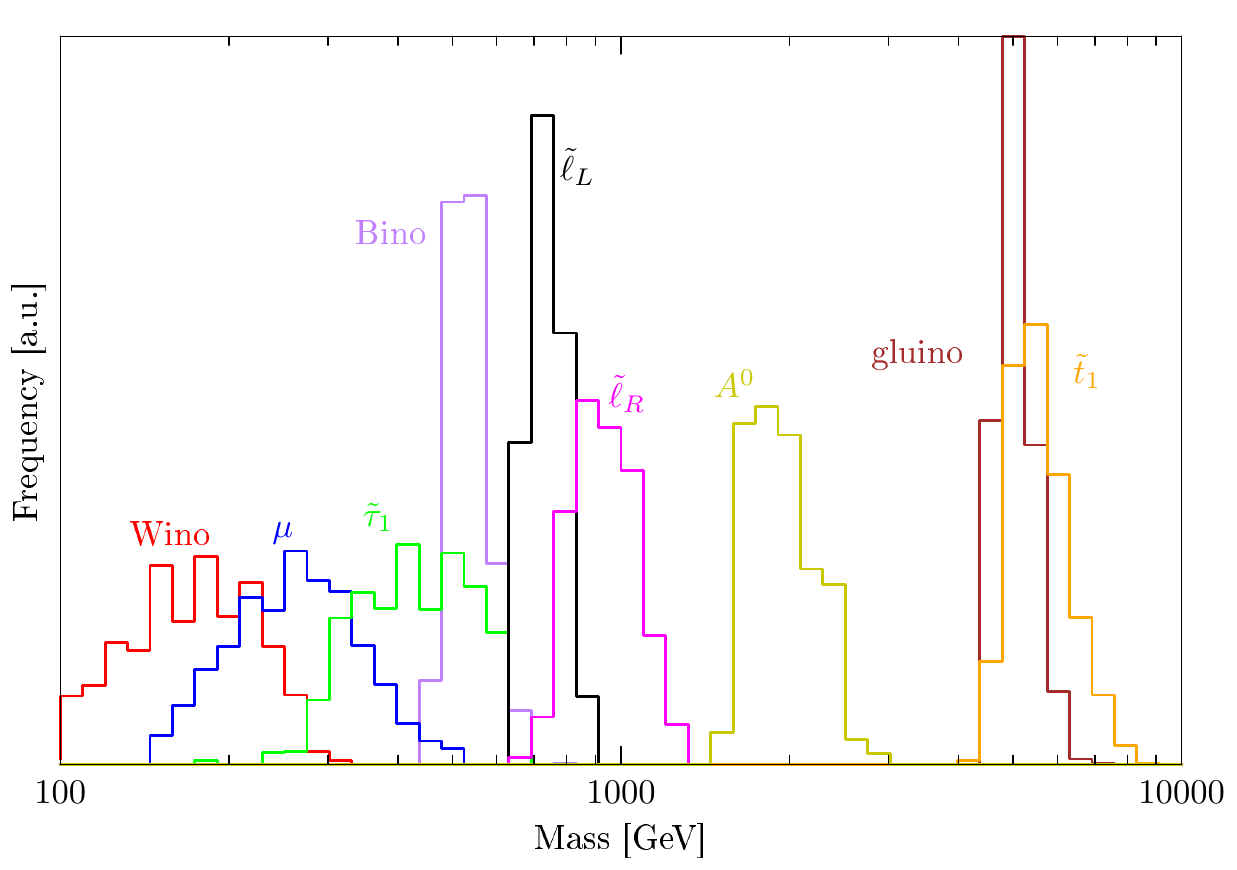}}
\caption{(a): Posterior distribution of the SUSY parameters consistent with $a_\mu$, the Higgs mass and vacuum stability bound.
We choose the parameter points at which $m_{h^0} > 124$ GeV and $a_\mu|_{\rm SUSY} > 19.2 \times 10^{-10}$.
(b): Posterior distribution with the LHC and LEP constraints 
(see Sec.~\ref{sec:LHC}).
In both cases, the predicted $\tan\beta$ is typically $\sim 30-40$. 
}
\label{fig:scan}
\end{figure}

In Fig.~\ref{fig:scan}, we present the result of the parameter scan of the present model.
We adopt log-flat priors for $M_{\rm mess}, M_{*}$, and linear-flat priors for $\Lambda_{\rm GMSB}^{L,D}$, $\Delta m_{H_u}^2$, $\Delta m_{H_d}^2$, imposing $M_* > 10^{11}$\,GeV and $M_{\rm mess} > 10^{6}$\,GeV.
We show the MSSM parameters which is  consistent with $a_\mu$, the Higgs mass and the vacuum stability bound on the stau direction~\cite{Hisano:2010re}.
The model predicts light Higgsinos and the Winos. 
The impact on the electroweak precision measurements are 
found to be minor~\cite{Cho:2011rk}.

\subsection{Effects of gravity mediated SUSY breaking}
As the Sweet Spot Supersymmetry assumes a rather large gravitino mass of $\order{1}$\,GeV, the gravity mediated SUSY breaking effects could cause 
FCNC/LFV and CP problems even if they are subdominant.
First, let us consider the gravity mediated contributions to the $B$-term~\cite{Moroi:2011fi}.
In supergravity, the 
$\mu$-term from the K\"ahler potential in Eq.\,\eqref{eq:Kmu} is shifted to
\begin{align}
\mu \simeq  \mu_0- \frac{ A_{Z_D}}{\Lambda_\mu} \times m_{3/2}\ ,
\end{align}
with $\mu_0 = F_D/\Lambda_\mu$ in Eq.\,\eqref{eq:mu}.
The associated $B$-term induced by the gravity mediation effects is
 \begin{align}
 \label{eq:BmuSUGRA}
    B \mu \simeq \mu_0 m_{3/2} +  2 \frac{A_{Z_D}}{\Lambda_\mu} m_{3/2}^2\ ,
 \end{align}
 where we have neglected the small contribution to $B\mu$ in Eq.\,\eqref{eq:Bmu}.
Note that $B\neq m_{3/2}$ due to the coupling between the holomorphic
term, $A_{Z_D} H_u H_d/\Lambda_\mu$,
in the K\"ahler potential and 
$W = w_D^2 Z_{D}+w_L^2 Z_{L}$ (see e.g., Ref.\,\cite{Evans:2013lpa}).%
\footnote{We assume that the effective K\"ahler potentials appear in the Einstein frame.
Even if they appear in the conformal frame, 
$B\mu/\mu \neq m_{3/2}$, and hence,
we obtain the similar 
effective CP phase in Eq.\,\eqref{eq:CPeff}. }
Those effects do not induce CP violating phases, 
since we have set $m_{3/2} > 0 $.

A CP violating phase appears from 
the additional origin of the $\mu$-term,
\begin{align}
    W = c_\mu\frac{w_D^2}{M_\mathrm{Pl}} H_u H_d\ ,
\end{align}
with $c_\mu$ being a complex valued coupling constant of $\order{1}$. 
This term is consistent with the PQ symmetry which is identified with the U(1)$_D$ symmetry.
This term shifts the $\mu$-term and $B$-term by,
\begin{align}
    &{\mit\Delta}\mu = -c_\mu\frac{w_D^2}{M_\mathrm{Pl}} =-\sqrt{3}c_\mu  \kappa_D m_{3/2}\ ,\\
    & {\mit\Delta} B \mu =-c_\mu\frac{w_D^2}{M_\mathrm{Pl}}  m_{3/2} = -\sqrt{3}c_\mu \kappa_D m_{3/2}^2  \ ,
\end{align}
where ${\mit\Delta} B \mu/{\mit\Delta}\mu =m_{3/2}$.
Thus, we find that the phases of the total $\mu$-term and the $B\mu$-term are no more aligned due to $c_\mu\neq 0$, which induces a CP violating phase.

The CP violating phase is estimated as follows.
In the phase convention in
the previous section, 
the phase of the total $\mu$-term induced by $c_\mu$ is 
\begin{align}
    \delta_{\mu} \sim \frac{
    {\mit \Delta}\mu}{\mu_0} \sim  \kappa_D \frac{m_{3/2}}{\mu_0}\times \mathrm{Arg}{(c_\mu)} \ .
\end{align}
Thus, in the phase convention where
the total $\mu$-term, $\mu_{\mathrm{tot}}>0$, 
the total $B\mu$-parameter at the scale $M_*$ becomes 
\begin{align}
    B \mu_{\mathrm{tot}} \simeq \mu_\mathrm{tot}m_{3/2} + 3 \frac{A_{Z_D} }{\Lambda_\mu} m_{3/2}^2e^{-i \delta_\mu}\ ,
\end{align}
where the first term is real-positive.
At the TeV scale, the $B\mu$-term is dominated by the gaugino mass contributions through the RG running, which is real valued in the present model.
As a result, the $B\mu$-term 
at the TeV scale is given by,
\begin{align}
B \mu \simeq \frac{m_A^2}{\tan\beta} + 3 \frac{A_{Z_D} }{\Lambda_\mu} m_{3/2}^2e^{-i \delta_\mu} \ ,
\end{align}
where $m_A$ is the mass of the CP-odd Higgs.
As a result, the effective CP-violating phase appearing in the $B\mu$-term is of 
\begin{align}
\label{eq:CPeff}
    \delta_{\mathrm{eff}} \sim 3\frac{A_{Z_D}}{\Lambda_\mu}\frac{\tan\beta m_{3/2}^2}{m_A^2} \delta_{\mu} 
    \sim 
    \frac{\tan\beta\kappa_D\Lambda_{D}^2}{\Lambda_\mu M_\mathrm{Pl}} \frac{m_{3/2}^3}{\mu_0 m_A^2}\times \arg(c_\mu)\ .
\end{align}
Thus, we find that the CP violating phase due to the gravity mediated effects on the $B\mu$-term is suppressed by $\order{10^{-9}}\times (m_{3/2}/\mathrm{GeV})^3$.
Thus, the expected electron EDM from Eq.\,\eqref{eq:EDM} is 
much lower than the current limit for $m_{3/2}\lesssim 10$\,GeV.

Next, let us consider the effects of 
the subdominant gravity mediated soft masses squared of the sfermions,
\begin{align}
 K \sim c_{ij} \frac{Z_{D,L}^\dagger Z_{D,L}}{M_\mathrm{Pl}^2} \phi_i^\dagger \phi_j \ ,
\end{align}
where $\phi$'s are MSSM matter chiral fields and $c_{ij}$ 
the $\order{1}$ coefficients.
In general, they are 
not flavor diagonal and have CP violating phases.

The CP violating sleptons squared mass matrix contributes to the electron EDM, which roughly 
correlates with the Bino contributions to $a_\mu|_\mathrm{SUSY}$ (see e.g., Ref~\cite{Feng:2001sq}) as,
\begin{align}
\left|\frac{d_e}{e}\right| &\sim \frac{m_e m_\tau}{m_\mu^3}
|\operatorname{Im}[\delta_{13}^{LL}
\delta_{13}^{RR}]|\times
a_\mu|_{\mathrm{Bino}} \ .
\end{align}
Here, $m_\tau$ is the tau lepton mass, $\delta_{13}^{LL} \sim m_{3/2}^2/m_{\tilde{\ell}}^2$
and $\delta_{13}^{LL} \sim m_{3/2}^2/m_{\tilde{e}_R}^2$ 
with $m_{\tilde{\ell}}^2$ and $m_{\tilde{e}_R}^2$ being the left-handed and the right-handed slepton squared masses.
The left-right mixing parameters are suppressed
since neither GMSB nor gravity mediation generates 
large $A$-terms.
As a result, the electron EDM is roughly given by,
\begin{align}
\left|\frac{d_e}{e}\right| \lesssim 4\times 10^{-34}\,\mathrm{cm}\times \left(\frac{a_\mu|_\mathrm{SUSY}}{2\times 10^{-9}}\right)
\left(\frac{m_{3/2}}{1\,\mathrm{GeV}}\right)^4
\left(
\frac{500\,\mathrm{GeV}}{m_{\tilde{\ell},\tilde{e}_R}}
\right)^{4}
\times |\arg[\delta_{13}^{LL} \delta_{13}^{RR}]|
\ ,
\end{align}
where we used $a_\mu|_\mathrm{Bino}\lesssim a_\mu|_\mathrm{SUSY}$.
Thus, the gravity mediated contribution to the electron EDM 
through the slepton mass is consistent with the current upper limit on the electron EDM 
for $m_{3/2} = \order{1}$\,GeV.

The flavor violation in the slepton soft masses also induce the LFV processes.%
\footnote{The FCNC in the quark sector is highly suppressed for $m_{3/2}=\order{1}$\,GeV by squark masses in the TeV range~\cite{Gabbiani:1996hi}.}
For example, the branching ratio of $\mu\to e+\gamma$ is roughly correlated with $a_\mu|_\mathrm{SUSY}$ (see e.g. Ref.\,\cite{Feng:2001sq}) as,
\begin{align}
\mathrm{Br}(\mu\to e+\gamma) 
&\sim 
\frac{12\pi^2}{G_F^2m_\mu^4}\,
(\delta_{12}^{LL,RR})^2\times a_\mu|_{\mathrm{SUSY}}{}^2  \ , \\
& \sim 10^{-18} \times \left(
\frac{a_\mu|_\mathrm{SUSY}}{2\times 10^{-9}}\right)^2
\left(\frac{m_{3/2}}{1\,\mathrm{GeV}}\right)^4
\left(
\frac{500\,\mathrm{GeV}}{m_{\tilde{\ell},\tilde{e}_R}}
\right)^{4}
\ ,
\end{align}
where $G_F$ is the Fermi constant.
Thus, for $a_\mu|_\mathrm{SUSY} = \order{10^{-9}}$,
the induced branching ratio is much smaller than 
the current upper limit by MEG experiment~\cite{TheMEG:2016wtm},
\begin{align}
    \mathrm{Br}(\mu\to e+\gamma) < 4.2 \times 10^{-13}\ ,
\end{align}
for $m_{3/2}=\order{1}$\,GeV.

The other LFV processes, $\mu\to3e$ 
and $\mu\to e$ conversion, are also correlated with $\mu\to e+\gamma$ for large $\tan\beta$, 
where both of them are dominated by the contributions of the Penguin diagrams.
Roughly, they are $\mathrm{Br}(\mu\to 3e) 
\sim \alpha \times
\mathrm{Br}(\mu\to e + \gamma)$ and 
$C_R(\mu\to e) \sim Z\alpha/\pi \times \mathrm{Br}(\mu\to e + \gamma)$, respectively (see e.g. \cite{Hisano:1995cp,Calibbi:2006nq}).%
\footnote{$C_R$ denotes the conversion rate in a nucleus divided by the muon capture rate by a nucleus.} 
Here, $\alpha$ is the fine-structure constant and $Z$ is the atomic number in a nucleus. 
The predicted rates are far below the current upper limits, $\mathrm{Br}(\mu\to 3e) < 1.0\times 10^{-12}$ by SINDRUM I 
\cite{Bellgardt:1987du}
and $C_R(\mu\to e\mbox{ in Au})<7\times 10^{-13}$ 
by SINDRUM II experiment~\cite{Bertl:2006up}, respectively. 

If the gravitino mass is no much less than $10$\,GeV, the predicted
electron EDM and the LFV 
are within the reach of the future experiments.
Those include the further improvement of the EDM measurements~\cite{Andreev:2018ayy,PhysRevLett.103.223001}, 
$Br(\mu\to e+\gamma) < 6.0\times 10^{-14}$
(MEG-II \cite{Baldini:2013ke}), 
$\mathrm{Br}(\mu\to 3 e) \lesssim 10^{-16}$ (Mu3e \cite{Blondel:2013ia}) and 
$C_R(\mu \to e \mbox{ in Al})\lesssim 3\times 10^{-17}$ (Mu2e \cite{Bartoszek:2014mya}, COMMET \cite{Adamov:2018vin}).

Finally, let us comment on the subdominant gravity mediation contribution to the gaugino masses and the trilinear $A$-term.
If the SUSY breaking fields are completely singlets under any symmetries, we expect the gravity mediated effects on those parameters of $\order{m_{3/2}}$.
In our setup, however, the SUSY breaking fields are charged under U(1)$_{D,L}$ and $R$-symmetry. 
Accordingly, the gravity mediated effects are of $\order{m_{3/2}^3/M_\mathrm{Pl}^2}$, which are negligible.
Besides, the anomaly mediated contributions~\cite{Randall:1998uk,Giudice:1998xp}
are aligned with GMSB, since $m_{3/2}$ in the superpotential is taken to be real positive.
Therefore, there are no SUSY CP
problems from the gravity/anomaly mediated contributions to the gaugino masses and the $A$-terms.

\section{LHC Signatures}
\label{sec:LHC}
Here we discuss the LHC constraints on the present model.
To achieve large $a_\mu|_\mathrm{SUSY}$, the masses of the relevant SUSY particles are rather small, which suffer from the LHC constraints~\cite{Abdughani:2019wai, Chakraborti:2020vjp, Chakraborti:2021kkr, Endo:2020mqz, Hagiwara:2017lse}. 
In the present GMSB models, the lightest SUSY particle (LSP) is the gravitino and, all the MSSM particles can decay into the goldstini.
For example, partial decay rate of the Wino into the goldstino/gravitino is given by
\begin{align}
\frac{c}{\Gamma(\tilde{W} \to \tilde G_{L}  W)}\sim 2\times 10^{13}\,{\mathrm{m}} \times\kappa_L^2\left(\frac{m_{3/2}}{1~{\rm GeV}} \right)^2\left(\frac{m}{100~{\rm GeV}} \right)^{-5}.
\end{align}
Unless the gravitino is much lighter than $\order{1}$\,MeV, the MSSM particles cannot decay inside the LHC detector.%
\footnote{The relic abundance of the NLSP is severely constrained by the Big-Bang Nucleosynthesis (BBN)
when its lifetime is longer than $\order{10^2}$\,sec~\cite{Shirai:2010rr}.
In the present model, the NLSP abundance depends on 
the cosmological evolution of the pseudo-flat directions~\cite{Ibe:2006rc}, which will be discussed in future work.}

\subsection{Higgsino and Wino system}

In the present model, so-called the GUT relation among the gaugino masses are violated.
In the typical parameter region of interest, the next-to-lightest SUSY particle (NLSP) is the Wino or Higgsino.%
\footnote{Strictly speaking, the NLSP is the massive goldstino in our model. In the following, however, we call the lightest SUSY particle in the MSSM sector the NLSP as in the conventional context of the GMSB phenomenology.}
The other particles are heavier than these particles and play less important roles at the LHC, compared to the Wino and Higgsinos.
Therefore, a simplified setup of the Higgsino-Wino system is useful  to see the collider constraints on the present model.

The collider signature significantly depends on the nature of the NLSP.

\subsubsection*{Wino NLSP}
After the electroweak symmetry breaking, the Wino particles are decomposed into a neutralino  $\tilde {\chi}^0_1$ and a chargino $\tilde {\chi}^\pm_1$.
The tree-level mass difference between the charged Wino and neutral Wino is approximately given by
\begin{align}
     m_{\tilde {\chi}^\pm_1} -  m_{\tilde {\chi}^0_1} |_{\rm Wino-like} 
     = \frac{m_W^4  \sin^2(2\beta) t_W^2} {\mu^2 (M_1 - M_2)} 
     + \frac{m_W^4 \cos^2(2\beta) M_2}{2 \mu^4} + \cdots\ ,
\end{align}
with $M_1$, $\mu > M_2 > 0$ \cite{Hall:2012zp}.
For a sizable $a_\mu|_{\mathrm{SUSY}}$ , $\tan\beta$ should be large and accordingly, $\sin(2\beta) \simeq 2/\tan\beta$ is suppressed.
Therefore the tree-level mass splitting is severely suppressed. 
In addition to the tree-level mass splitting, the electroweak loop correction provides the mass difference around 165\,MeV \cite{Ibe:2012sx}.
In the present model, if the Higgsino mass is greater than 300\,GeV, the Wino mass splitting is less than 1\,GeV and the charged Wino can be long-lived and the decay length can be $\order{1}$\,cm.
This charged tracks are detected as a disappearing charged track at colliders.
This signature is intensively studied in the anomaly mediation model \cite{Ibe:2006de, Buckley:2009kv, Asai:2007sw, Asai:2008sk, Asai:2008im, Mahbubani:2017gjh, Fukuda:2017jmk, Saito:2019rtg, Chigusa:2019zae}.
The latest ATLAS search of the disappearing charged tracks with 139 fb$^{-1}$ data excludes the Wino lighter than 660\,GeV for a large Higgsino mass \cite{ATLAS-CONF-2021-015}.

\subsubsection*{Higgsino NLSP}
By the electroweak symmetry breaking, the Higgsinos are decomposed into two neutralinos  $\tilde {\chi}^0_{1,2}$ and a chargino $\tilde {\chi}^\pm_1$.
As in the case of the Wino, there are mass splittings among these particles.
The mass difference between the charged Higgsino and lightest neutral Higgsino is approximately given as\cite{Nagata:2014wma,Fukuda:2017jmk,Fukuda:2019kbp},
\begin{align}
   m_{\tilde {\chi}^\pm_1} -  m_{\tilde {\chi}^0_1}|_{\rm Higgsino-like} =
    \frac{m_W^2}{2}\left(  \frac{1}{M_2}  +\frac{t_W^2}{M_1} \right)- \sin(2\beta) 
    \frac{m_W^2}{2}\left(  \frac{1}{M_2}  -\frac{t_W^2}{M_1}   \right) + \cdots.
\end{align}
The mass difference between the two neutralinos is approximately twice of this mass splitting.
If the Wino mass is less than 1\,TeV, the chargino-neutralino mass splitting is greater than 5\,GeV.
Therefore we do not expect the disappearing charged tracks like the Wino NLSP case.
There is, however, another important signature comes from the decay of the $\tilde {\chi}^0_{2} \to \tilde {\chi}^0_{1} \ell^+ \ell^-$.
This soft lepton signature can give the Higgsino mass constraint up to 200\,GeV~\cite{Aad:2019qnd}.

\subsubsection*{Next-to-NLSP decay}
In addition to the constraint on the direct production of the NLSP, the decay of the next-to-NLSP into the NLSP also provides the clue for the collider searches.
If the Higgsino mass is much greater than the Wino mass, the Higgsinos decay into the Wino with $Z, W^{\pm}$ and $h^0$ bosons with almost equal branching fractions.
Such bosons decay into leptons, photons and $b$-jets, which are characteristic signatures.

In the present analysis, we study the following analysis.
\begin{description}
    \item [Disappearing charged track] \mbox{}\\
 We directly apply the ATLAS result of the Wino LSP searches with  139\,fb$^{-1}$ data  \cite{ATLAS-CONF-2021-015}.
 
    \item [Soft di-lepton]\mbox{}\\
We directly adopt the Higgsino LSP search with 139\,fb$^{-1}$ data  at the ATLAS \cite{Aad:2019qnd}.

    \item [Tri-lepton mode from $W$ and $Z{(h^0)}$ decays]\mbox{}\\
We study the search of leptonic decays of $W^{(*)}$ and $Z^{(*)}$ with  139\,fb$^{-1}$ data  at the ATLAS \cite{ATLAS-CONF-2020-015}.
We use model-independent inclusive event selections (12 on-shell channels and 17 off-shell channels).

    \item [One lepton + two b-jets from $W$ and $h^0$ decays]\mbox{}\\
    We study the search of a leptonic $W$ decay and $b$-jets from the 125 GeV Higgs boson with  139\,fb$^{-1}$ data at the ATLAS  \cite{Aad:2019vvf}.%
    \footnote{Other search channels such as $h^0\to \gamma\gamma$~\cite{Aad:2020qnn} 
    are less important.}
    
    We adopt the signal regions, SR-LM, SR-MM and SR-HM.
    In our estimation, we found this search cannot give a constraint on the present Higgsino-Wino system.
    
    \item [Di-lepton from two $W(\tilde{\ell})$ bosons decay]\mbox{}\\
         We study the search of two leptonic $W$ decays or two slepton decays into a lepton and a dark matter with  
         139\,fb$^{-1}$ data at the ATLAS     \cite{Aad:2019vnb}.
     We adopt eight signal regions with di-lepton.
         In our estimation, this search cannot give a constraint on the present Higgsino-Wino system.
    \item [Mono-jet]\mbox{}\\
    The ATLAS search of mono-jet events \cite{Aad:2021egl} aims the direct production of the dark matter with high energy initial state radiations.
    The constraint of this search on the Wino is comparable to the LEP chargino searches \cite{LEP}, if we adopt the leading order cross section.

\end{description}
In our analysis, we have used the programs MadGraph5\_MC@NLO \cite{Alwall:2011uj,Alwall:2014hca}, PYTHIA8 \cite{Sjostrand:2014zea} and DELPHES 3 \cite{deFavereau:2013fsa}  (with FastJet  \cite{Cacciari:2011ma} incorporated).
We adopt the cross sections provided by the LHC SUSY Cross Section Working Group \cite{crosssection}, which are based on works
~\cite{Bozzi:2007qr,Debove:2010kf,Fuks:2012qx,Fuks:2013vua,Fuks:2013lya,Fiaschi:2018xdm,Beenakker:1999xh,Fiaschi:2018hgm}.
If constraints on the simplified model provided by the LHC is directly applicable to the present model, we recast the constraints.

In addition to the direct production of the SUSY particles, Wino and Higgsinos affect the SM processes.
Precision measurement of the Drell–Yan process can give indirect signature of the SUSY particles at the LHC \cite{Matsumoto:2017vfu, Matsumoto:2018ioi,Katayose:2020one,DiLuzio:2018jwd}.
Although such searches can also provide the constraint on the low mass Wino, we need to know the detailed information on the systematic uncertainty of the measurement and the SM signature estimation at the LHC.
At present, we cannot get a reliable constraint from the precision measurement at the LHC.

\begin{figure}[t]
\centering
	\includegraphics[width=0.5\hsize,clip]{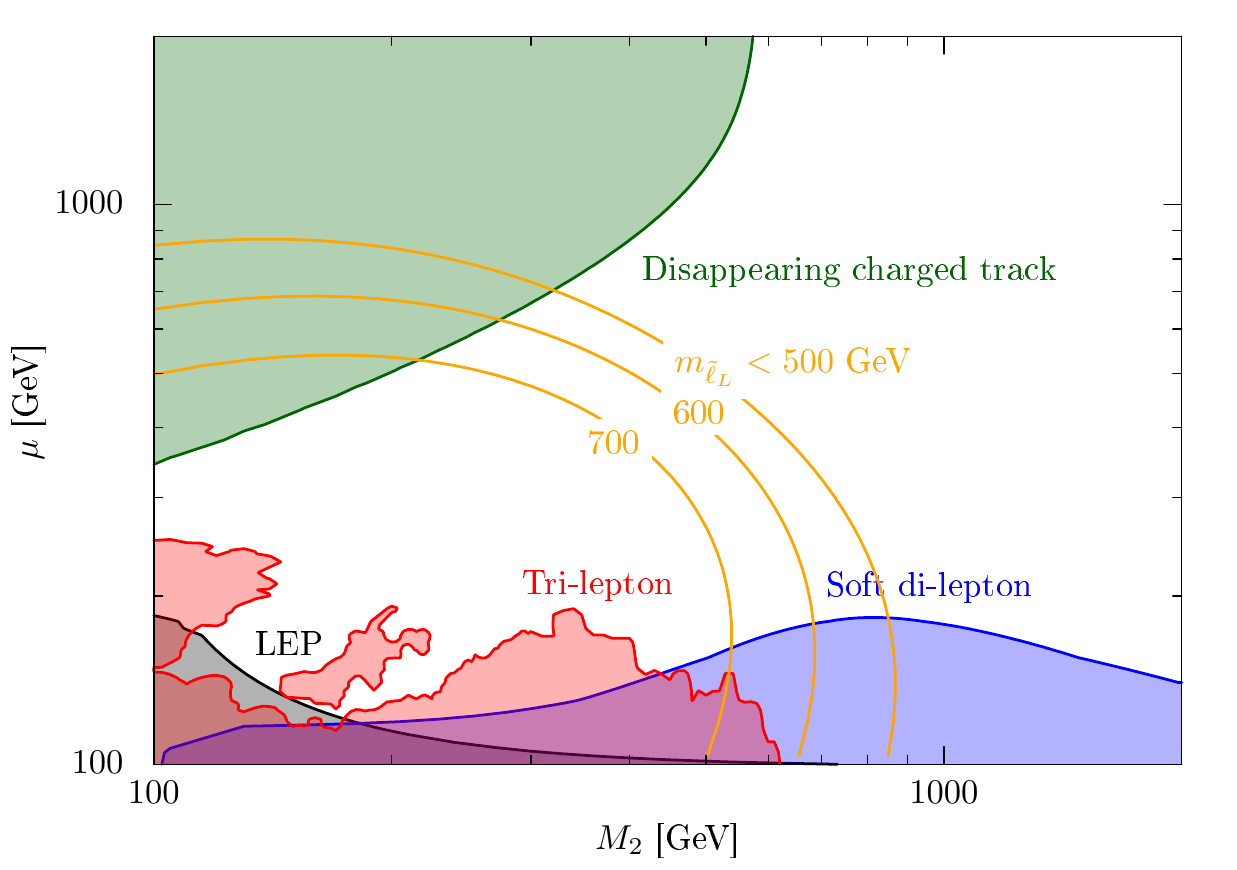}
	\caption{The current collider constraints on the Higgsino-Wino system.
	The orange lines 
	show the rough upper limits on the left-handed slepton mass to explain the observed $a_\mu$.
	}
	\label{fig:LHC}
\end{figure}

In Fig.~\ref{fig:LHC}, we show the current LHC and LEP chargino constraints on the plane of the Wino and Higgsino mass parameters $M_2$ and $\mu$ with $\tan\beta = 40$.
We show the excluded regions by the disappearing charged track in green, soft di-lepton in blue and tri-lepton in red.

In the present model, 
the dominant contributions to $a_\mu|_{\mathrm{SUSY}}$ come from the left-handed slepton-Wino-Higgsino loops.
Thus, for given $M_2$, $\mu$ and $\tan\beta$, we 
can predict the left-handed slepton mass
to explain $a_\mu$. In Fig.~\ref{fig:LHC}, we also show the
rough upper limit on the left-handed slepton mass to explain the observed $a_\mu$ by 
the Higgsino-Wino contribution at the one-loop level.
The figure shows that the LHC constraint, $m_{\tilde{\ell}_L} > 660$\,GeV, favors the Higgsino-Wino mass within $100$\,GeV--$600$\,GeV.

\subsection{Scalar lepton constraint}
To explain  $a_\mu$, the scalar leptons should also be light.
In the present model, the left-handed slepton is typically lighter than the right-handed slepton  since $\Lambda_\mathrm{GMSB}^L \ll \Lambda_\mathrm{GMSB}^D$.
Therefore the constraint on the left-handed sleptons is relevant for the present model, 
where the left-handed sleptons dominantly decay into the Wino-like chargino and neutralino.

If the Wino is the NLSP, we can directly apply the constraint on simplified model of $\tilde{\ell}_L \to \ell \tilde{\chi}^0_1\,(\ell =e, \mu)$  provided by the ATLAS \cite{Aad:2019vnb}.
In this model, the ATLAS searches for the di-leptons from the process $pp \to \tilde{\ell}^+
\tilde{\ell}^-
\to \ell^+ \ell^- \tilde{\chi}^0_1 \tilde{\chi}^0_1 $ are relevant.
If the mass of $\tilde{\chi}^0_1$ is less than around 300\,GeV, the current upper bound on the cross section of the slepton pair production in the simplified model is around 0.3\,fb.

In the case of $\tilde{\ell}_{L}, \tilde{\nu}$ and Wino system, the sneutrino production also contributes the di-lepton signature, as the sneutrino can decay into the charged Wino with a charged lepton.
The branching fractions are $\mathrm{BF}(\tilde{\ell}^{-} \to \tilde W^{-} \nu)  = 2 \mathrm{BF}(\tilde{\ell}^{-} \to \tilde W^{0} \ell^{-})$,
and 
$\mathrm{BF}(\tilde{\nu} \to \tilde W^{+} \ell^{-})  = 2 \mathrm{BF}(\tilde{\nu} \to \tilde W^{0} \nu)$.
Therefore, the constraint of the left-handed slepton can be obtained by the condition that%
\footnote{Here, we do not distinguish $\tilde{W}^\pm$ from $\tilde{W}^0$ to apply the analysis in Ref.~\cite{Aad:2019vnb}, as the decay products of the $\tilde{W}^\pm$ is too soft to affect the LHC di-lepton analysis. }
\begin{align}
    \frac{1}{9} \sigma(\tilde{\ell}^+ \tilde{\ell}^-  )
+    \frac{4}{9} \sigma(\tilde{\nu} \tilde{\nu} )    
+    \frac{2}{9} \sigma(\tilde{\ell}^{\pm} \tilde{\nu} )     = 0.3\,\mathrm{fb}\ ,
\end{align}
where the prefactors represent the branching fractions to the di-lepton modes.
As a result, we find the lower limit on the left-handed slepton mass, $m_{\tilde{\ell}_L} > 660$\,GeV, in the Wino NLSP case.

\subsection{Heavy Higgs constraint}
In the present model, the $B$ term is zero at the UV scale and radiatively generated through the RG effects (Fig.~\ref{fig:RGE}).
Therefore the heavier Higgs mass tends to be light (Fig.~\ref{fig:all_pdf}).
Moreover to enhance the muon $g-2$, the value of the $\tan\beta$ is large.
In this case, the production of the heavier CP-odd Higgs $A^0$ is significantly enhanced and can be constrained by the LHC experiments.
The CMS and ATLAS provide constraints on the process $pp \to A^0 \to \tau^+ \tau^-$ \cite{Sirunyan:2018zut,Aad:2019byo} and these constraints have a significant impact on the present model.
In this model, there are SUSY particles lighter than the CP-odd Higgs mass, and hence, $A^0$ can also decay into such SUSY particles.
For the large $\tan\beta$, however, the branching fractions into the SUSY particles are small and we directly apply the constraint on the $m_{A^0}-\tan\beta$ provided by the ATLAS \cite{Aad:2019byo}.

In Fig.~\ref{fig:collider_pdf}, we show the posterior distribution of the SUSY parameters after imposing the collider constraints.
In our estimation so far, the light Wino and Higgsino of masses $100$--$300$\,GeV are consistent with the LHC constraints.
Compared to the constraints on simplified models studied by the CMS and ATLAS, our constraints look rather conservative.
One reason of such weak constraints will be 
the mass degeneracy 
of the Higgsino and Wino with which the SUSY events at the LHC are less energetic.
Moreover, in the present model, the Higgsino and Wino have various decay channels.
Therefore, the current LHC searches optimized for simplified models are not so effective.
For the typical masses of the Higgsinos and Wino in the present model, the production cross section is rather large.
Thus, if we can optimize the LHC searches for the present model, we have a large chance of the discovery of the extended Sweet Spot Supersymmetry for muon $g-2$.
Study of such prospects are beyond the focus of this paper.
We will discuss this point in the future work.

Let us comment on the case of $\tilde {\tau}$ NLSP case.
In this case, the stau is long-lived, and the LHC signatures are massive charged tracks.
The direct constraint on the stau mass is 430\,GeV \cite{Aaboud:2019trc}.
In this stau NLSP case, a portion of the stau is stopped in the LHC detectors.
By measuring the late-time decay of the stopped stau into goldstini \cite{Asai:2009ka,Ito:2011xs}, we can obtain the fundamental information on the SUSY breaking sectors \cite{Buchmuller:2004rq}.

In the present model, the Wino and Higgsino are light $\sim 300$ GeV.
Although the direct production of such particles will be out of reach of the ILC250, such light particles have significant impact on the SM processes through quantum corrections.
Therefore, the precision measurement of the di-fermion process $e^+e^- \to f \bar{f}$ can probe the most of the parameter space consistent with the muon $g-2$ \cite{Harigaya:2015yaa}.
The slepton mass is, on the other hand, predicted to be relatively high and it will be difficult to probe the slepton directly at the ILC500 \cite{Baum:2020gjj}.

\section{Conclusions}
In this paper, we discussed the GMSB models which explain $a_\mu$ and the Higgs boson mass simultaneously.
There have been two known major types of gauge-mediated models that explain the observed mass of the Higgs boson and the $a_\mu$ of the muon. 
The first type is a model that produces a large $A$-term by mixing the Higgs fields with messenger fields. 
This class of the models predict
the existence of relatively light squarks and gluinos. 
Therefore, in those models, it is expected that particles with color charges are produced at the LHC, which puts severe collider constraints.
In the second type of model, the mass of the Higgs boson is realized by heavy squarks, which evades severe LHC constraints.
As we have discussed, however, the
GUT relation of the GMSB should be violated
since the explanation of $a_\mu$ requires light sleptons.
The naive violation of the GUT relation ends up with the SUSY CP problem.
To avoid such problems, we proposed a model in which the phases of the gaugino masses are aligned despite the violation of the GUT relation.

The successful explanation
of $a_\mu$ and the Higgs boson mass requires a rather light $\mu$-parameter and 
heavy squarks.
To achieve such a SUSY spectrum, we need additional sources of the Higgs soft masses squared other than the GMSB contributions.
The model also requires the origin of the $\mu$-term which is free from the CP violation.
For these purposes, 
we utilized the (extended) Sweet Spot Supersymmetry~\cite{Ibe:2007km}.
As we have shown, the 
model can explain $a_\mu$ and the Higgs boson mass in the GMSB model without causing the SUSY CP problem.
The model evades the LHC constraints so far.
We also found that the SUSY CP, FCNC, LFV
processes caused by the subdominant gravity mediation are also suppressed.

Several comments are in order.
In our set up, we only considered the Down-type and the Lepton-type messengers.
By utilizing the product GUT models, 
it is also possible to have GUT violating  messenger fields of more various representations, such as the adjoint representation (see e.g., 
Refs.~\cite{Ibe:2012qu,Yanagida:2017dao}).
The extended Sweet Spot Supersymmetry allows those complicated messenger sector without causing unwanted CP violation.

Since we assume $\Lambda_{D,L} \simeq 10^{15\mbox{--}16}$\,GeV, 
the masses of the pseudo-flat directions 
in Eq.\,\eqref{eq:Zmass} are in the hundreds GeV 
to a few TeV region for $m_{3/2}=\order{1}$\,GeV.
The cosmological evolution of a light pseudo-flat direction has been discussed in Ref.\,\cite{Ibe:2006rc}.
Note that there are two pseudo-flat directions in the  present model. 
Besides, there is a goldstino with a mass $2m_{3/2}$
in addition to the gravitino.
Since the massive goldstino has 
a cosmological lifetime~\cite{Cheung:2010mc},
cosmic ray signatures of the very late time decay of the goldstino could give us a smoking gun of the present model.
We will discuss details of cosmology of the model including the dynamics of the pseudo-flat directions, the constraints for the gravitino/goldstino dark matter in future work.

\section*{Acknowledgments}
This work is supported by Grant-in-Aid for Scientific Research from the Ministry of Education, Culture, Sports, Science, and Technology (MEXT), Japan, 17H02878 (M.I. and S.S.), 18H05542 (M.I.), 18K13535, 19H04609, 20H01895, 20H05860 and 21H00067 (S.S.), and by World Premier International Research Center Initiative (WPI), MEXT, Japan. 
This work is also supported by the Advanced Leading Graduate Course for Photon Science (S.K.), the JSPS Research Fellowships for Young Scientists (S.K. and Y.N.) and International Graduate Program for Excellence in Earth-Space Science (Y.N.).

\appendix
\section{Higgs Coupling to SUSY Breaking}
\label{sec:HiggsSSS}
In Sec.~\ref{sec:SSS}, we consider the effective K\"ahler potential which generates the $\mu$-term 
as well as the additional soft masses squared of the Higgs doublets.
In this appendix, we discuss an example of the UV completion~\cite{Kitano:2006wm,Kitano:2006wz,Ibe:2006rc,Ibe:2007km}.
The simplest example is based on the O'Raifeartaigh model with the superpotential,
\begin{align}
\label{eq:Higgs}
W = w^2 Z + \frac{1}{2}\lambda Z X^2 + M_{XY} XY
+ h H_u \bar{q} X + \bar{h} H_d q X + M_q q\bar{q}
+ m_{3/2}M_\mathrm{Pl}^2\ .
\end{align}
Here, $Z$, $X$, $Y$ are gauge singlet fields, 
$H_{u,d}$ are the Higgs doublets, and $(q,\bar{q})$
are the vector-like
SU(2)$_L$ doublet fields.
All the phases of the coupling constants, $\lambda$, $h$, and $\bar{h}$, as well as those of the mass parameters $M_{XY}$ and $M_{q}$ can be rotated away
without loss of generality.
We assume the PQ symmetry with the charges, $PQ(H_{u,d})=1$, $PQ(X)=-1$,
$PQ(Y)=1$ and $PQ(Z)=2$, which is explicitly broken by $w^2$.
In the Sweet Spot Supersymmetry in Sec.~\ref{sec:SSS}, we identified $Z$ in Eq.\,\eqref{eq:Higgs} with $Z_D$.

\begin{center}
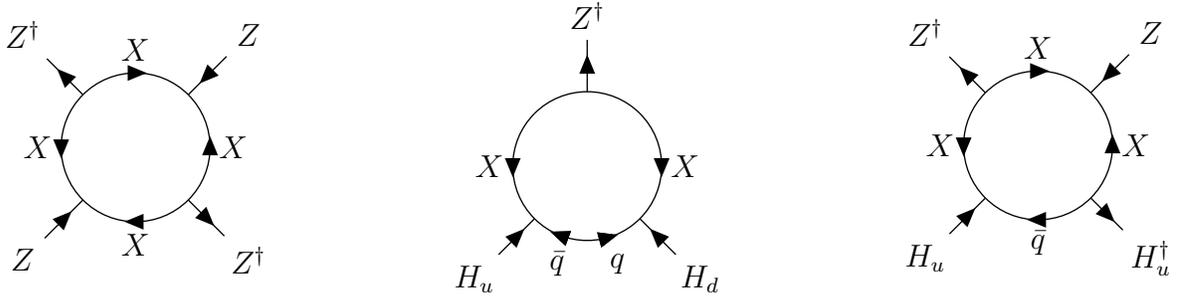
\begin{figure}[tpb]  
\begin{minipage}{.33\linewidth}
\begin{center}      
\begin{tikzpicture} 
\begin{feynhand}    
\vertex [particle] (z2) at (1.5,1.5) {$Z$};
\vertex [particle] (z1) at (-1.5,1.5) {$Z^\dagger$};
\vertex [particle] (z3) at (-1.5,-1.5) {$Z$};
\vertex [particle] (z4) at (1.5,-1.5) {$Z^\dagger$};
\vertex [dot] (d1) at (-0.7,0.7);
\vertex [dot] (d2) at (0.7,0.7);
\vertex [dot] (d3) at (-0.7,-0.7);
\vertex [dot] (d4) at (0.7,-0.7);
\propag[fer] (d1) to [quarter left](d2);
\propag[fer] (d1) to [quarter right](d3);
\propag[fer] (d4) to [quarter left](d3);
\propag[fer] (d4) to [quarter right](d2);
\propag[fer] (d1) to (z1);
\propag[fer] (z2) to (d2);
\propag[fer] (z3) to (d3);
\propag[fer] (d4) to (z4);
\vertex [particle] (x1) at (1.3,0) {$X$};
\vertex [particle] (x2) at (-1.3,0) {$X$};
\vertex [particle] (x3) at (0,1.3) {$X$};
\vertex [particle] (x4) at (0,-1.3) {$X$};
\end{feynhand}
\end{tikzpicture}
\end{center}
\end{minipage}
\begin{minipage}{.33\linewidth}
\begin{center}      
\begin{tikzpicture} 
\begin{feynhand}    
\vertex [particle] (z1) at (0,2) {$Z^\dagger$};
\vertex [particle] (z3) at (-1.5,-1.5) {$H_u$};
\vertex [particle] (z4) at (1.5,-1.5) {$H_d$};
\vertex [dot] (d1) at (-0.7,0.7);
\vertex [dot] (d2) at (0.7,0.7);
\vertex [dot] (d3) at (-0.7,-0.7);
\vertex [dot] (d4) at (0.7,-0.7);
\vertex [dot] (d5) at (0,0.989949);
\propag[plain] (d1) to [quarter left](d2);
\propag[fer] (d1) to [quarter right](d3);
\propag[antmaj] (d4) to [quarter left](d3);
\propag[fer] (d2) to [quarter left](d4);
\propag[fer] (d5) to (z1);
\propag[fer] (z3) to (d3);
\propag[fer] (z4) to (d4);
\vertex [particle] (x1) at (1.3,0) {$X$};
\vertex [particle] (x2) at (-1.3,0) {$X$};
\vertex [particle] (x4) at (0.4,-1.3) {$q$};
\vertex [particle] (x4) at (-0.4,-1.3) {$\bar{q}$};
\end{feynhand}
\end{tikzpicture}
\end{center}
\end{minipage}
\begin{minipage}{.33\linewidth}
\begin{center}      
\begin{tikzpicture} 
\begin{feynhand}    
\vertex [particle] (z2) at (1.5,1.5) {$Z$};
\vertex [particle] (z1) at (-1.5,1.5) {$Z^\dagger$};
\vertex [particle] (z3) at (-1.5,-1.5) {$H_u$};
\vertex [particle] (z4) at (1.5,-1.5) {$H_u^\dagger$};
\vertex [dot] (d1) at (-0.7,0.7);
\vertex [dot] (d2) at (0.7,0.7);
\vertex [dot] (d3) at (-0.7,-0.7);
\vertex [dot] (d4) at (0.7,-0.7);
\propag[fer] (d1) to [quarter left](d2);
\propag[fer] (d1) to [quarter right](d3);
\propag[fer] (d4) to [quarter left](d3);
\propag[fer] (d4) to [quarter right](d2);
\propag[fer] (d1) to (z1);
\propag[fer] (z2) to (d2);
\propag[fer] (z3) to (d3);
\propag[fer] (d4) to (z4);
\vertex [particle] (x1) at (1.3,0) {$X$};
\vertex [particle] (x2) at (-1.3,0) {$X$};
\vertex [particle] (x3) at (0,1.3) {$X$};
\vertex [particle] (x4) at (0,-1.3) {$\bar{q}$};
\end{feynhand}
\end{tikzpicture}
\end{center}
\end{minipage}
\caption{The Feynman diagrams in the UV model which generate the Higgs mass parameters.}
\label{fig:Diagram}
\end{figure}
\end{center}

\begin{figure}[t]
\centering
	\includegraphics[width=0.5\hsize,clip]{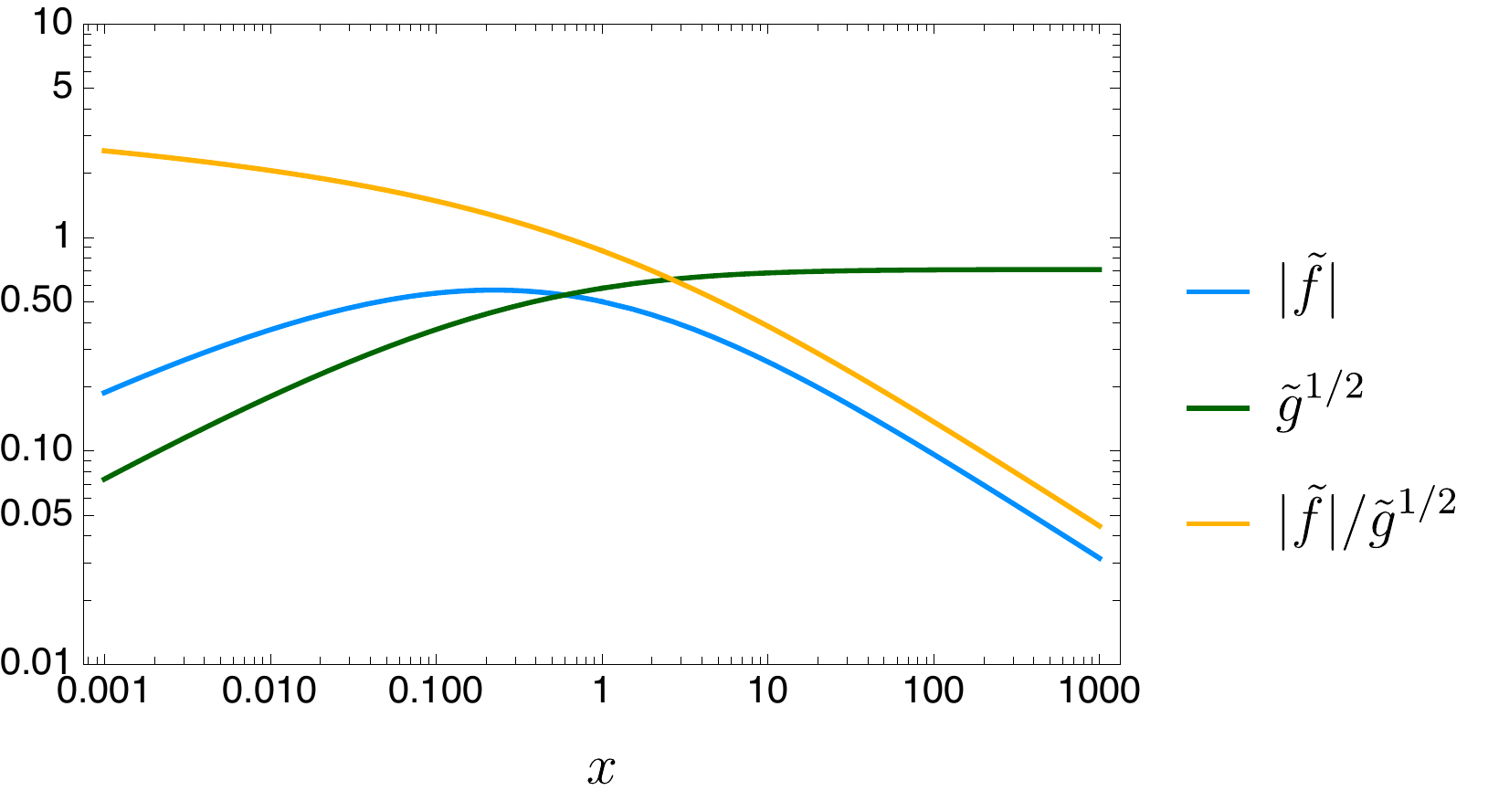}
	\caption{The functions $|\tilde{f}|$, $\tilde{g}$ 
	and their ratio $|\tilde{f}|/\tilde{g}^{1/2}$.}
	\label{fig:functions}
\end{figure}

The coefficient of the $|Z|^4$ term in the effective K\"ahler potential is given by,
\begin{align}
\label{eq:ZZZZ}
    \frac{1}{\Lambda^2} = \frac{1}{4} \lambda^4 \int \frac{d^4\ell_E}{(2\pi)^4}\frac{\ell_{E}^2}{(M_{XY}^2+\ell_E^2)^4}  = \frac{\lambda^4}{12(4\pi)^2 M_{XY}^2}\ .
\end{align}
Here, $\ell_E$ is the Euclidean loop momentum.
The coefficient of $Z^\dagger H_u H_d$ term is given by,
\begin{align}
\label{eq:ZHH}
&\frac{1}{\Lambda_\mu} = \int \frac{d^4\ell_E}{(2\pi)^4}
\frac{1}{(M_{XY}^2+\ell_E^2)^2}\frac{M_{q}}{(M_{q}^2+\ell_E^2)} = - \frac{\lambda h \bar{h}}{(4\pi)^2 M_{XY}}\cdot \tilde{f}\left(\frac{M_{XY}^2}{M_q^2}\right) \ ,
\end{align}
with
\begin{align}
\label{eq:f}
\tilde{f}(x) = x^{1/2}\times\frac{1-x+\log x}{(1-x)^2}\ .
\end{align}
The coefficient of $Z^\dagger Z H_{u} H_{u}$ term is given by,
\begin{align}
\label{eq:ZZHH}
    &\frac{1}{\Lambda_u^2} = \lambda^2 h^2 \int \frac{d^4\ell_E}{(2\pi)^4}\frac{1}{(M_{XY}^2+\ell_E^2)^3}\frac{\ell_{E}^2}{(M_{q}^2+\ell_E^2)} = \frac{\lambda^2h^2}{(4\pi)^2}\frac{1}{M_{XY}^2}
    \cdot\tilde{g}\left(\frac{M_{XY}^2}{M_q^2}\right)\ , 
\end{align}
with
\begin{align}
\label{eq:g}
    \tilde{g}(x) = x\times\frac{-3 +4 x -x^2 -2\log x}{2(1-x)^3} \ .
\end{align}
The coefficient of $Z^\dagger Z H_d^\dagger H_d$ is given by replacing $h$ with $\bar{h}$.
In Fig.\,\ref{fig:functions}, we show the function $\tilde{f}$ and $\tilde{g}$.
As is clear from the integrands of Eqs.\eqref{eq:ZZZZ}, \eqref{eq:ZHH} and \eqref{eq:ZZHH}, the 
integration is dominated at the loop momentum of  $\order{M_{XY}}$, and hence, the scale $M_*$ at which 
the Higgs mass parameters are generated 
is given by $M_* \simeq M_{XY}$.

From Eqs.\,\eqref{eq:ZHH} and \eqref{eq:ZZHH}, 
we find that 
\begin{align}
    \frac{\Lambda_u}{\Lambda_\mu} = \frac{h}{(4\pi)}\frac{|\tilde{f}|}{\tilde{g}^{1/2}}\ ,
\end{align}
where the ratio $|\tilde{f}|/\tilde{g}^{1/2}$ is of $\order{1}$ in a wide range of $M_{XY}^2/M_q$.
Thus, to provide the appropriate $\Lambda_\mu$ and $\Lambda_{u,d}$ in 
Eqs.\,\eqref{eq:Lambdamu} and \eqref{eq:Lambdaud}, 
we find that $h$ and $\bar{h}$ are of $\order{1}$. With this choice,  the generated $\mu$-parameter is parametrically smaller than the additional Higgs soft term by an order of magnitude.

Note also that $Z$ obtains non-vanishing $A$-term VEV due to the supergravity effect,
\begin{align}
    \langle Z\rangle = \frac{\sqrt{3}\Lambda^2}{6M_\mathrm{Pl}} \ .
\end{align}
In order for $ \langle Z\rangle$ not to affect the O'Raifeartaigh model, we require $\lambda \langle Z \rangle < M_{XY}$
\begin{align}
\label{eq:lambda}
    {32\sqrt{3}\pi^2}\frac{M_{XY}}{M_\mathrm{Pl}} < \lambda^3
    \ .
\end{align}
If $Z$ is an independent of the GMSB, $M_{XY}$ (that is $M_*$) is a free parameter as long as $\lambda$ satisfies Eq.\,\eqref{eq:lambda}.
Instead, if we identify $Z$ with 
either $Z_{D}$ or $Z_{L}$, $M_*$
should be at around the sweet spot in Eq.\,\eqref{eq:sweetspotMstar}
as discussed in Sec.\,\ref{sec:SSS}.

\bibliographystyle{apsrev4-1}
\bibliography{references}
\end{document}